\documentclass{conm-p-l}

\copyrightinfo{2006}{}

\usepackage{graphicx}
\usepackage{subfigure}

\usepackage{amssymb,amsthm,amscd,amsmath}

\usepackage{pstricks}
\usepackage{pst-plot}

\usepackage{latexsym}
\usepackage{tikz} 
\usetikzlibrary{arrows,decorations.markings}

\newtheorem{theorem}{Theorem}[section]

\theoremstyle{definition}

\theoremstyle{remark}

\numberwithin{equation}{section}

\newcommand{\tu}{\tilde{u}}
\newcommand{\hu}{\hat{u}}

\newcommand{\e}{\epsilon}

\newcommand{\ga}{\gamma}

\newcommand{\Dl}{\Delta}
\renewcommand{\th}{\theta}
\newcommand{\ra}{\rightarrow}
\newcommand{\al}{\alpha}
\newcommand{\be}{\beta}
\newcommand{\sg}{\sigma}

\newcommand{\pa}{\partial}

\newcommand{\tG}{\tilde{G}}
\newcommand{\om}{\omega}
\newcommand{\Om}{\Omega}

\renewcommand{\O}{{\mathcal O}}

\newcommand{\htau}{\hat{\tau}}
\newcommand{\hrho}{\hat{\rho}}

\newcommand{\vphi}{\varphi}

\newcommand{\non}{\nonumber}

\newcommand{\hth}{\hat{\theta}}

\begin{document}

\title[Chaotic Flux Dynamics]
{On the Chaotic Flux Dynamics in a Long Josephson Junction}

\author{Z. C. Feng}
\address{Department of Mechanical and Aerospace Engineering, 
University of Missouri, Columbia, MO 65211}
\email{fengf@missouri.edu}

\author{Y. Charles Li}
\address{Department of Mathematics, University of Missouri, 
Columbia, MO 65211}
\curraddr{}
\email{cli@math.missouri.edu}
\thanks{}

\subjclass{Primary 35, 82; Secondary 37}
\date{}

\dedicatory{}


\begin{abstract}
Flux dynamics in an annular long Josephson junction is studied. Three main topics 
are covered. The first is chaotic flux dynamics and its prediction via Melnikov 
integrals. It turns out that DC current bias cannot induce chaotic flux dynamics, 
while AC current bias can. The existence of a common root to the Melnikov integrals 
is a necessary condition for the existence of chaotic flux dynamics. The second 
topic is on the components of the global attractor and the bifurcation in the 
perturbation parameter measuring the strength of loss, bias and irregularity of the 
junction. The global attractor can contain co-existing local attractors e.g. a local 
chaotic attractor and a local regular attractor. In the infinite dimensional phase 
space setting, the bifurcation is very complicated. Chaotic attractors can appear and 
disappear in a random fashion. Three types of attractors (chaos, breather, spatially 
uniform and temporally periodic attractor) are identified. The third topic is ratchet 
effect. Ratchet effect can be achieved by a current bias field which corresponds to 
an asymmetric potential, in which case the flux dynamics is ever lasting chaotic. 
When the current bias field corresponds to a symmetric potential, the flux dynamics 
is often transiently chaotic, in which case the ratchet effect disappears after 
sufficiently long time. 
\end{abstract}

\maketitle

\tableofcontents








\section{Introduction}

The recently developed theory \cite{Li04} on chaos in partial differential 
equations has the greatest potential of significance 
in its abundant applications in science and engineering. The 
variety of the specific problems also stimulates innovation of the theory. 
In these representative publications \cite{Li03a} \cite{Li03b} \cite{Li04c} 
\cite{LM94} \cite{LMSW96} \cite{Li99} \cite{Li04a} \cite{Li04b}, two categories of the theory 
were developed. The category developed in \cite{Li03a} \cite{Li03b} \cite{Li04c} involves transversal 
homoclinic orbits, and a shadowing technique is used to prove the existence of 
chaos. This category is very complete. The category in \cite{LM94} \cite{LMSW96} 
\cite{Li99} \cite{Li04a} \cite{Li04b} deals with Silnikov (non-transversal) homoclinic orbits, 
and a geometric construction 
of Smale horseshoes is employed. This category is not very complete. The main machineries for 
locating homoclinic orbits are (1). Darboux transformations, (2). Isospectral theory, 
(3). Persistence of invariant manifolds and Fenichel fibers, (4). Melnikov analysis and shooting 
technique. Overall, the two categories of the theory on chaos in partial differential equations
can be regarded as a new development at the intersection among Integrable Theory, Dynamical 
System Theory, and Partial Differential Equations \cite{Li04}. In this article, we will apply 
the above chaos theory to study the chaotic flux dynamics in a long Josephson junction.

A Josephson junction consists of three parts: two superconductors separated by a thin ($\sim 10$ \AA ) 
dielectric barrier. The main character of a Josephson junction is that it behaves as a single 
superconductor. In particular, electric flux called fluxons can travel through the dielectric 
barrier through Josephson tunneling. In general, the scenario is as follows: when two superconductors 
are separated by a macroscopic distance, their phases can change independently. As the two superconductors
are moved closer to about $30$ \AA \ separation, quasiparticles can flow from one superconductor to the 
other by means of single electron tunneling. When the separation is reduced to $10$ \AA , Cooper 
pairs can flow from one superconductor to the other (Josephson tunneling). In this case, phase 
correlation is realized between the two superconductors, and the whole Josephson junction behaves as a single 
superconductor. This phenomenon was predicted by Brian Josephson in 1962 \cite{Jos74}. It has 
significant applications in quantum-mechanical circuits. The governing equations are \cite{BP82}
\[
V(t) = \frac{\hbar}{2e} \frac{d u}{d t}, \quad I(t) = I_c \sin u(t)
\]
where $V(t)$ and $I(t)$ are the voltage and current across the Josephson junction, $u$ is the 
phase difference between the wave functions in the two superconductors, the constant $I_c$ is called 
the critical current, and the constant $\frac{\hbar}{2e}$ is the magnetic flux quantum. Interesting 
simple phenomena can be observed from the above governing equations, e.g. when $V(t)$ is constant 
$V_0$ in $t$, $u$ is linear in $t$, then the current $I(t)$ will be an AC current with amplitude 
$I_c$ and frequency $\frac{2e}{\hbar}V_0$. This shows that a Josephson junction can be a perfect 
voltage-to-frequency converter. Significant applications of a Josephson junction can be found in 
many areas, e.g. in medicine for measurement of small currents in the brain and the heart. Josephson 
junction may also provide key ingredients for future quantum computers. For more details on the physics 
of the Josephson junction, see \cite{BP82}. 

When a Josephson junction has one or more dimensions longer than the rest, i.e. a long 
Josephson junction (LJJ), the phase difference $u$ is also a function of spatial coordinates 
along the longer dimensions. In 
this article, we will focus on one longer dimension case. It is well known that the flux dynamics in a long
Josephson junction (LJJ) is described by the so-called perturbed sine-Gordon equation \cite{BP82}
\begin{equation}
u_{tt} = c^2 u_{xx} +\sin u +\e f 
\label{PSG}
\end{equation}
where again $u$ is the phase difference between the two superconductors, and $c$ is a constant. 
The above equation is the rescaled standard form. In dealing with real junctions, one must take into 
account losses, bias, and junction irregularities which influence the flux dynamics 
\cite{BP82}. These effects are of perturbation nature, and accounted for by the term $\e f$ where $\e$ is 
the perturbation parameter. 
Different forms of $\e f$ can even by set up in experiments. Typically $\e f$ takes the following forms:
\begin{itemize}
\item $\e f = -\e \al u_t +\e \ga$, where $-\e \al u_t$ represents shunt loss and $\e \ga$ 
represents DC current bias \cite{BP82}. 
\item $\e f = -\e \al u_t +\e \be u_{txx} +\e \ga$, where $\e \be u_{txx}$ represents longitudinal loss
\cite{BP82}.
\item $\e f = -\e \al u_t +\e \ga + \e a \sin \om t$, where $\e a \sin \om t$ represents AC current 
bias \cite{BFU04}. 
\item $\e f = -\e \al u_t +\e \ga \cos x + \e a \sin \om t$, where $\e \ga \cos x$ represents AC 
current field \cite{Bec05} \cite{SK04}.
\item $\e f = -\e \al u_t +\e g(x) + \e a \sin \om t$, where $\e g(x)$ represents spatially periodic  
current field \cite{Bec05}. 
\end{itemize}
Of course, many other forms of $\e f$ can also be set up in experiments. There is an abundant literature
on long Josephson junctions, for a sample, see \cite{Kie09, Aug09, Boy08, Tor07, Sob06, Rau04, Pan02, Kas01}.

Travelling wave solutions of (\ref{PSG}) satisfy an ordinary differential equation which had been 
studied numerically since as early as 1968 by Johnson \cite{Joh68} using an analog-digital computer, 
till most recently \cite{BGV03} via analytical and numerical tools.

In this article, we shall study the full partial differential equation. When $\e =0$, equation 
(\ref{PSG}) is the well-known sine-Gordon equation. Two well-known types of solutions to the 
sine-Gordon equation are the kink and breather solutions. The simplest kink solution is 
independent of the space variable $x$. Such a simple kink can be observed in experiemnts for 
Josephson junctions. As $t \ra \pm \infty$, the kink $u(t)$ approaches e.g. $0$ and $2\pi$. In 
terms of the current and voltage ($I(t),V(t)$) mentioned above, this 
corresponds to a closed loop --- named a vortex or a fluxon. In the infinite dimensional phase 
space setting of the current article, this simple kink also represents a heteroclinic orbit. A 
breather oscillates periodically in time (breathing) and decays in space ($x \ra \pm \infty$). 
In our current setting, we switch the space and time, therefore, the breather is now spatially 
periodic and temporally homoclinic, i.e. a homoclinic orbit. For long Josephson junctions, 
spatial dependence is significant, dynamics beyond the simple kink becomes crucial. More 
sophisticated solutions incorporating both the kink and the breather have been constructed 
\cite{Li04e}. In the phase space, these represent heteroclinic orbits which depend on both time 
and space. Together, these heteroclinic orbits form a two-dimensional heteroclinic cycle. The 
current article will focus its study on the neighborhood of this heteroclinic cycle. Three main 
topics will be covered. The first is chaotic flux dynamics and its prediction via Melnikov 
integrals (sections 4,5,6). The second is on the components of the global attractor and the 
bifurcation in the perturbation parameter (section 7). The third is ratchet effect (section 8). 
Section 9 is the conclusion.

\section{The Phase Space}

We pose periodic boundary condition on the above perturbed sine-Gordon equation (\ref{PSG}),
\begin{equation}
u(t,x+2\pi ) = u(t,x).
\label{pbc}
\end{equation}
This boundary condition corresponds to an annular LJJ \cite{SK04} \cite{Bec05}. Convenient for 
mathematical studies, a more restricted condition --- the even constraint may be imposed 
\begin{equation}
u(t,-x)=u(t,x)
\label{ec}
\end{equation}
which corresponds to a mirror symmetry of the annular LJJ. We restrict the parameter $c$ to the interval 
$c \in (0.5,1)$ to minimize the number of unstable modes of $u=0$ to two. (\ref{PSG}) is invariant 
under $u \ra u+2\pi$. When $\e =0$, it is also invariant under $u \ra -u$. (\ref{PSG}) is globally 
well-posed \cite{Li04c}, i.e. for any $(u^0,u^0_t) \in H^{n+1}\times H^n$ (the Sobolev spaces on 
$[0,2\pi ]$) where $n \geq 1$ is an integer, there exists a unique mild solution to (\ref{PSG}),
$(u(t),u_t(t)) \in C([0, \infty ), H^{n+1}\times H^n)$, such that $(u(0),u_t(0)) = (u^0,u^0_t)$. 
One can introduce the evolution operator $F^t$ as $(u(t),u_t(t)) = F^t (u^0,u^0_t)$. For any fixed $t \in 
[0, \infty )$, $F^t$  is a $C^\infty$ map. The space $H^{n+1}\times H^n$ will be the phase space where 
the perturbed sine-Gordon flow (\ref{PSG}) is defined. 

\section{Isospectral Integrable Theory}

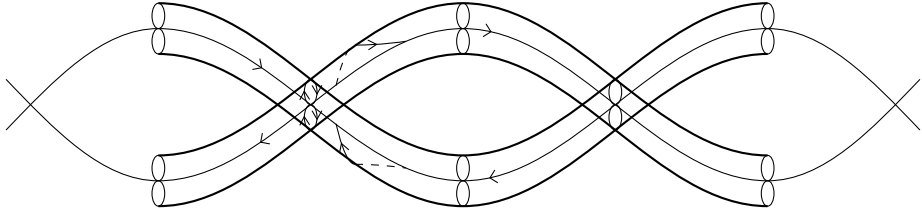
\begin{figure}[ht]
\vspace{0.5in}
\begin{tikzpicture}[scale=1.35] 
  \draw [postaction={decorate, decoration={markings, mark=at position 4cm with {\arrow{angle 90}}, mark=at position 7.5cm with {\arrowreversed{angle 90}} }}] (0,-.25) sin (1.5,.75) cos (3,0) sin (4.5,-.75) cos (6,0) sin (7.5,.75) cos (9,-.25) ; 
  \draw[thick] (1.5,.5) cos (3,-.25) sin (4.5,-1) cos (6,-.25) sin (7.5,.5); 
  \draw[thick] (1.5,1) cos (3,.25) sin (4.5,-.5) cos (6, .25) sin (7.5,1); 
  \draw (1.5,.625) ellipse (.0625 and .125); 
  \draw (1.5,.875) ellipse (.0625 and .125); 
  \draw (7.5,.875) ellipse (.0625 and .125); 
  \draw (7.5,.625) ellipse (.0625 and .125); 
  \draw [postaction={decorate, decoration={markings, mark=at position 4cm with {\arrowreversed{angle 90}}, mark=at position 7.5cm with {\arrow[black]{angle 90}} }}] (0,.25) sin (1.5,-.75) cos (3,0) sin (4.5,.75) cos (6,0) sin (7.5,-.75) cos (9,.25); 
  \draw[thick] (1.5,-.5) cos (3,.25) sin (4.5,1) cos (6,.25) sin (7.5,-.5); 
  \draw[thick] (1.5,-1) cos (3,-.25) sin (4.5,.5) cos (6,-.25) sin (7.5,-1); 
  \draw (1.5,-.625) ellipse (.0625 and .125); 
  \draw (1.5, -.875) ellipse (.0625 and .125); 
  \draw (7.5,-.875) ellipse (.0625 and .125); 
  \draw (7.5,-.625) ellipse (.0625 and .125); 
\draw [postaction={decorate, decoration={markings, mark=at position .003cm with {\arrowreversed{angle 60}}, mark=at position .425cm with {\arrowreversed{angle 60}} }}] (3,.125) ellipse (.0625 and .125); 
\draw [postaction={decorate, decoration={markings, mark=at position .003cm with {\arrowreversed{angle 60}}, mark=at position .425cm with {\arrowreversed{angle 60}} }}] (3,-.125) ellipse (.0625 and .125); 
\draw (6,.125) ellipse (.0625 and .125); 
\draw (6,-.125) ellipse (.0625 and .125); 
\draw (4.5,-.625) ellipse (.0625 and .125); 
\draw (4.5,-.875) ellipse (.0625 and .125); 
\draw (4.5,.625) ellipse (.0625 and .125); 
\draw (4.5,.875) ellipse (.0625 and .125); 
\draw[dashed] (3.25,.1875) parabola[bend at end] (3.4375,.5875); 
\draw[postaction={decorate, decoration={markings, mark=at position .3cm with {\arrow{angle 60}} }}] (3.4375,.5875) parabola[bend at start] (3.9375,.625); 
\draw[postaction={decorate, decoration={markings, mark=at position .3cm with {\arrowreversed{angle 60}} }}] (3.25,-.1875) parabola[bend at end] (3.4375,-.5875); 
\draw[dashed] (3.4375,-.5875) parabola[bend at start] (3.9375,-.625); 
\end{tikzpicture} 
\caption{A topological illustration of the heteroclinic cycle.}
\label{topi}
\end{figure}

When $\e =0$, the unstable manifolds of $u_+ = 0$ and $u_-=2\pi$, $W^u_\pm$ form a 2-dimensional 
heteroclinic cycle whose explicit expression can be obtained through Darboux transformations 
\cite{Li04e}:
\begin{eqnarray}
u_1 &=& 2\arccos (\tanh \tau ) \non \\
    & & -4 \arctan \bigg [ \frac{\sg ( \pm \tanh \tau \ \mbox{sech}\  \htau 
\cos x - \ \mbox{sech}\  \tau \tanh \htau )}{1-\sg ( \tanh \tau 
\tanh \htau \pm \ \mbox{sech}\  \tau \ \mbox{sech}\  \htau \cos x )}
\bigg ]\ , \label{cyc1} \\
u_2 &=& 2 \pi - 2\arccos (\tanh \tau ) \non \\
    & & -4 \arctan \bigg [ \frac{\sg ( \pm \tanh \tau \ \mbox{sech}\  \htau 
\cos x + \ \mbox{sech}\  \tau \tanh \htau )}{1-\sg ( \tanh \tau 
\tanh \htau \mp \ \mbox{sech}\  \tau \ \mbox{sech}\  \htau \cos x )}
\bigg ]\ , \label{cyc2} 
\end{eqnarray}
where the ranges of $\arccos$ and $\arctan$ are [$0,\pi$] and ($-\pi /2,
\pi /2$), and ($\tau , \sg , \htau$) are given by
\[
\tau = t-\rho , \ \sg =\sqrt{1-c^2}, \  \htau = \sg t -\hrho = \sg \tau - \Dl \rho , \ 
\Dl \rho = \hrho - \sg \rho ,
\]
and $\rho$ and $\hrho$ are two real parameters. This heteroclinic cycle satisfies both 
the periodic boundary condition and the even constraint (\ref{pbc}) and (\ref{ec}). A topological illustration of 
the heteroclinic cycle is shown in Figure \ref{topi}. The heteroclinic cycle incorporates both kink and breather 
characteristics. The homoclinic degeneracy to be discussed later corresponds to the breather characteristic 
(when time and space are switched). The current article will focus its study near this heteroclinic cycle.

Via isospectral theory, Melnikov vectors can be derived for the sine-Gordon equation \cite{Li04e}. 
Along each of $u_1$ and $u_2$ in the above heteroclinic cycle, explicit expressions of the 
Melnikov vectors can be obtained through Darboux transformations \cite{Li04e}:
\begin{eqnarray}
\frac{\pa F_0}{\pa u_t} &=& \frac{64 \pi}{c^4}(\tG_1/\phi_1+\tG_2/\phi_2)(\tG_3/\phi_1+\tG_4/\phi_2)\ ,
\label{MV1} \\
\frac{\pa F_1}{\pa u_t} &=& \frac{\pi (1-c^2)}{4c}i (\nu_1^2-\nu_0^2)\frac{\overline{\Phi_1}
\overline{\Phi_2}}{(\nu_1 |\Phi_1|^2 +
\overline{\nu_1}|\Phi_2|^2)(\overline{\nu_1} |\Phi_1|^2 +\nu_1 |\Phi_2|^2)}\ ,
\label{MV2}
\end{eqnarray}
where $F_0$ and $F_1$ are two invariants of the sine-Gordon equation, $\nu_0 = i/4$, 
$\nu_1 =\frac{1}{4}[c +i\sqrt{1-c^2}]$,
\begin{equation}
\phi = 2 \left ( \begin{array}{c} \cosh \frac{\tau }{2} \cos \frac{\th }{2}
-i \sinh \frac{\tau }{2} \sin \frac{\th }{2} \cr 
-\cosh \frac{\tau }{2} \sin \frac{\th }{2}
-i \sinh \frac{\tau }{2} \cos \frac{\th }{2} \cr
\end{array}\right )\ ,
\label{phi}
\end{equation}
here $\th = \pi /2$ for $u_1$ and $\th = -\pi /2$ for $u_2$;
\begin{equation}
\Phi = \left ( \begin{array}{lr} -\nu_0 \phi_2/\phi_1 & \nu_1 \cr 
-\nu_1 & \nu_0 \phi_1/\phi_2 \cr
\end{array}\right ) \vphi \ , 
\label{bphi}
\end{equation}
and 
\[
\vphi = 2 \left ( \begin{array}{c} \cosh \frac{\htau }{2} \cos \frac{\xi }{2}
-i \sinh \frac{\htau }{2} \sin \frac{\xi }{2} \cr 
-\cosh \frac{\htau }{2} \sin \frac{\xi }{2}
-i \sinh \frac{\htau }{2} \cos \frac{\xi }{2} \cr
\end{array}\right )\ ,
\]
here $\xi = x +\hth$, $\hth = \pm \pi /2$, and the `$\pm$' corresponds to the `$\pm$' in $u_1$ 
and $u_2$; finally
\begin{eqnarray*}
\tG_1 &=& |\nu_1|^2 \frac{\nu_1 |\Phi_1|^2 +\overline{\nu_1}|\Phi_2|^2}
{\overline{\nu_1} |\Phi_1|^2 +\nu_1 |\Phi_2|^2} - \nu_0^2 \ , \\
\tG_2 &=& \frac{\nu_0 (\nu_1^2-\overline{\nu_1}^2) \Phi_1 \overline{\Phi_2}}
{\overline{\nu_1} |\Phi_1|^2 +\nu_1 |\Phi_2|^2} \ , \\
\tG_3 &=& \frac{\nu_0 (\nu_1^2-\overline{\nu_1}^2) \overline{\Phi_1}\Phi_2}
{\nu_1 |\Phi_1|^2 +\overline{\nu_1}|\Phi_2|^2} \ , \\
\tG_4 &=& |\nu_1|^2 \frac{\overline{\nu_1} |\Phi_1|^2 +\nu_1 |\Phi_2|^2}
{\nu_1 |\Phi_1|^2 +\overline{\nu_1}|\Phi_2|^2} - \nu_0^2 \ .
\end{eqnarray*}
The two Melnikov vectors $\frac{\pa F_0}{\pa u_t}$ and $\frac{\pa F_1}{\pa u_t}$ will be used to 
heuristically measure the splitting of the heteroclinic cycle when the sine-Gordon equation is 
under perturbations (\ref{PSG}).

\subsection{Homoclinic Degeneracy \label{HD}}

Inside the 2-dimensional unstable manifold $W^u_+$ of $u_+=0$, there is a homoclinic orbit 
asymptotic to $u_+=0$. Let $\rho \ra - \infty$ in the expression of $u_1$ (\ref{cyc1}), we get
\begin{equation}
u_1 \ra \mp 4 \arctan  \frac{\sg \ \mbox{sech}\  \htau 
\cos x}{1-\sg \tanh \htau } ,
\label{lhd}
\end{equation}
and $u_2 \ra 2\pi + u_1$. The other limit $\rho \ra + \infty$ produces equivalent expressions. 
Expression (\ref{lhd}) represents a pair of homoclinic orbits asymptotic to $u_+=0$ (cf: Figure 
\ref{topi}). One can also generate the expression (\ref{lhd}) directly from the Darboux transformation. 

In the limit of $\rho \ra - \infty$, the $\phi$ defined in (\ref{phi}) has the limiting property:
\[
\phi_2 / \phi_1 \ra - i .
\]
Thus the $\Phi$ defined in (\ref{bphi}) now has the form 
\begin{equation}
\Phi = \left ( \begin{array}{lr} -\frac{1}{4} & \nu_1 \cr 
-\nu_1 & -\frac{1}{4} \cr
\end{array}\right ) \vphi \ .
\label{nbp}
\end{equation}

In the limit of $\rho \ra - \infty$, the Melnikov vector $\frac{\pa F_0}{\pa u_t} \ra 0$ (\ref{MV1}), 
and the only Melnikov vector left is $\frac{\pa F_1}{\pa u_t}$ which is given by (\ref{MV2}) where 
$\Phi$ is now given by (\ref{nbp}). This Melnikov vector is transversal to the pair of homoclinic orbits
(\ref{lhd}). From Figure \ref{topi}, we see that the unstable and stable manifolds of $u_+=0$ intersect
along the pair of homoclinic orbits (\ref{lhd}). Inside the $\pa_x = 0$ plane, these two manifolds 
intersect transversally at $u_+=0$, thus we can use $\frac{\pa F_1}{\pa u_t}$ to heuristically measure the 
splitting of the homoclinic orbits when the sine-Gordon equation is under perturbations (\ref{PSG}). The 
rigorous mathematical problems are often open \cite{Li04e}.

\section{The Case of $f=-\al u_t +\be u_{txx} +\ga$}

\subsection{Heteroclinic Cycle Search}

In this case, we are interested in the possible heteroclinic cycle for the perturbed sine-Gordon equation 
(\ref{PSG}), connecting the two fixed points
\[
u_+ = -\arcsin (\e \ga ), \quad u_- = 2\pi-\arcsin (\e \ga ).
\]
Such a possible heteroclinic cycle has the potential of inducing chaos. Unfortunately, as shown later,
such a heteroclinic cycle does not exist. Nevertheless, we will present the failing arguments simply
because the argument in the current case is the simplest. Arguments in other cases are often similar.
Linearization at the fixed points leads to 
\[
u_{tt} = c^2 u_{xx} +\sqrt{1-(\e \ga )^2} u +\e [-\al u_t +\be u_{txx}].
\]
Let $u = \sum_{k=0}^\infty u_k(t) \cos kx$, then
\[
u''_k +\e (\al +\be k^2)u'_k +[c^2k^2 - \sqrt{1-(\e \ga )^2}]u_k = 0.
\]
Let $u_k \sim e^{\Om_k t}$, then
\[
\Om_k = -\frac{1}{2}\e (\al +\be k^2)\pm 
\sqrt{\sqrt{1-(\e \ga )^2}-c^2k^2+\frac{1}{4}\e^2 (\al +\be k^2)^2}.
\]
The results of \cite{Li04a} \cite{Li04e} imply the following invariant manifold theorem.
\begin{theorem}
When $c \in (1/2,1)$ and $\e \geq 0$ is sufficiently small, each of the two fixed points 
$u_\pm$ has a 2-dimensional $C^m$ ($m \geq 3$) unstable manifold $W^u_\pm$ and a 2-codimensional
$C^m$ center-stable manifold $W^{cs}_\pm$ in the phase space $(u,u_t) \in H^{n+1} \times H^n$ ($n \geq 1$). 
When $\e >0$, each of $u_\pm$ has a 2-codimensional $C^m$ local stable manifold $W^{s}_\pm$ of size 
$\O (\sqrt{\e})$. In a $\O (\sqrt{\e})$ neighborhood of $u_\pm$, $W^{s}_\pm = W^{cs}_\pm$. 
$W^u_\pm$ are $C^1$ in $\e$ for $\e \in [0, \e_0 )$ and some $\e_0 >0$. When $\be >0$, at 
$(u,u_t) \in H^{n+2} \times H^{n+2}$, $W^{cs}_\pm$ are $C^1$ in $\e$ for $\e \in [0, \e_0 )$.
When $\be =0$, $W^{cs}_\pm$ are always $C^1$ in $\e$ for $\e \in [0, \e_0 )$.
\end{theorem}
To the leading order, the signed distances (which are certain coordinate differences) between 
$W^u_\pm$ and $W^{cs}_\mp$ are given by the Melnikov integrals \cite{Li04a}
\begin{equation}
M_{j\ell }= \int_{-\infty}^{+\infty} \int_0^{2\pi} \left \{ \frac{\pa F_\ell}{\pa u_t} 
\left [-\al u_t +\be u_{txx} +\ga \right ] \right \} \bigg |_{u=u_j} dx dt, 
\label{MI}
\end{equation}
where ($j=1,2; \ \ell =0,1$) and $u_j$'s are given in (\ref{cyc1})-(\ref{cyc2}). 
The common zero of 
the Melnikov integrals and implicit function theorem 
imply the intersection between $W^u_\pm$ and $W^{cs}_\mp$. The intersected orbits approach $u_\pm$ 
as $t \ra -\infty$. As $t \ra +\infty$, they can reach the $\O (\sqrt{\e})$ neighborhood of 
$u_\mp$ where $W^{s}_\mp = W^{cs}_\mp$, as shown in \cite{Li04a}. Thus, they approach $u_\mp$ as 
$t \ra +\infty$, and form a heteroclinic cycle. So the key question now is whether or not the 
Melnikov integrals have a common zero. 
\[
M_{j\ell }= \al M_{j\ell }^{(\al )} + \be M_{j\ell }^{(\be )} +\ga M_{j\ell }^{(\ga )}
\]
where $M_{j\ell }^{(\cdot )}$ are functions of $c$ and $\Dl \rho = \hrho - \sg \rho$,
\begin{eqnarray*}
M_{j\ell }^{(\al )} &=& - \int_{-\infty}^{+\infty} \int_0^{2\pi} \left \{ \frac{\pa F_\ell}{\pa u_t}
u_t \right \} \bigg |_{u=u_j} dx dt, \\
M_{j\ell }^{(\be )} &=& \int_{-\infty}^{+\infty} \int_0^{2\pi} \left \{ \frac{\pa F_\ell}{\pa u_t}
u_{txx} \right \} \bigg |_{u=u_j} dx dt, \\
M_{j\ell }^{(\ga )} &=& \int_{-\infty}^{+\infty} \int_0^{2\pi} \left \{ \frac{\pa F_\ell}{\pa u_t}
 \right \} \bigg |_{u=u_j} dx dt.
\end{eqnarray*}
It turns out from numerical calculations (we believe it is analytically provable) that all the 
$M_{j\ell }^{(z)}$ ($z=\al , \be ,\ga ; j=1,2; \ell =0,1$) are real, and 
$M_{1\ell }^{(z)} = M_{2\ell }^{(z)}$ ($z=\al , \be ; \ell =0,1$) and $M_{1\ell }^{(\ga )} = -
M_{2\ell }^{(\ga )}$ ($\ell =0,1$). Therefore, the common zero of $M_{j\ell }$ satisfies the 
system
\begin{eqnarray}
& & \al M_{10}^{(\al )} + \be M_{10}^{(\be )} +\ga M_{10}^{(\ga )} = 0, \label{zm1} \\
& & \al M_{11}^{(\al )} + \be M_{11}^{(\be )} +\ga M_{11}^{(\ga )} = 0, \label{zm2} \\
& & \al M_{10}^{(\al )} + \be M_{10}^{(\be )} -\ga M_{10}^{(\ga )} = 0, \label{zm3} \\
& & \al M_{11}^{(\al )} + \be M_{11}^{(\be )} -\ga M_{11}^{(\ga )} = 0. \label{zm4}
\end{eqnarray}
There is no non-trivial solution to this system (\ref{zm1})-(\ref{zm4}), which shows a failure in 
searching for a heteroclinic cycle.  

\subsection{Heteroclinic Orbit Search \label{HC4}}

Next we try to search for an individual heteroclinic orbit. 
We only need to solve equations (\ref{zm1})-(\ref{zm2}) or (\ref{zm3})-(\ref{zm4}). In either case,
we obtain the equation
\[
\al =\chi \be , \quad \chi = \frac{M_{11}^{(\be )}M_{10}^{(\ga )}-M_{10}^{(\be )}M_{11}^{(\ga )}}
{M_{10}^{(\al )}M_{11}^{(\ga )}-M_{11}^{(\al )}M_{10}^{(\ga )}}.
\]
Since both $\al$ and $\be$ have to be positive, $\chi$ has to be positive too. But direct calculation 
shows that $\chi$ is always negative (see Table \ref{chi}). This shows a failure in searching 
for even an individual heteroclinic orbit. 

\begin{table}[ht]
$$\begin{array}{|c|c|c|c|c|c|}\hline
\Delta \rho & -10 & -5 & 0 & 5 & 10\\ \hline
c=0.55 & -1.0000 & -1.1617 & -1.4179 & -1.1617 & -1.0000\\ \hline
c=0.65 & -0.9998 & -1.0895 & -1.5753 & -1.0895 & -0.9998 \\ \hline
c=0.75 & -0.9997 & -1.0423 & -1.7368 & -1.0423 & -0.9997 \\ \hline
c=0.85 & -0.9990 & -1.0075 & -1.9272 & -1.0075 & -0.9990 \\ \hline
c=0.95 & -0.9875 & -0.9870 & -2.0338 & -0.9870 & -0.9875 \\ \hline
\end{array}$$
\caption{The table of $\chi$ for the heteroclinic orbit case in section \ref{HC4}.}
\label{chi}
\end{table}

\subsection{Homoclinic Orbit Search}

To search the homoclinic orbit asymptotic to $u_+=0$, we need to calculate the Melnikov integral:
\begin{equation}
M = \int_{-\infty}^{+\infty} \int_0^{2\pi} \left \{ \frac{\pa F_1}{\pa u_t} 
\left [-\al u_t +\be u_{txx} +\ga \right ] \right \} \bigg |_{u=u_1} dx dt, 
\label{HMI1}
\end{equation}
where $u_1$ is given in (\ref{lhd}), $\frac{\pa F_1}{\pa u_t}$ is given by (\ref{MV2}) where 
$\Phi$ is now given by (\ref{nbp}). 
\[
M = \al M^{(\al )} + \be M^{(\be )} + \ga M^{(\ga )} ,
\]
where 
\begin{eqnarray*}
M^{(\al )} &=& - \int_{-\infty}^{+\infty} \int_0^{2\pi} \left \{ \frac{\pa F_1}{\pa u_t}
u_t \right \} \bigg |_{u=u_1} dx dt, \\
M^{(\be )} &=& \int_{-\infty}^{+\infty} \int_0^{2\pi} \left \{ \frac{\pa F_1}{\pa u_t}
u_{txx} \right \} \bigg |_{u=u_1} dx dt, \\
M^{(\ga )} &=& \int_{-\infty}^{+\infty} \int_0^{2\pi} \left \{ \frac{\pa F_1}{\pa u_t}
 \right \} \bigg |_{u=u_1} dx dt.
\end{eqnarray*}
Numerical calculations show that $M^{(\ga )}$ is zero, 
$M^{(\al )}$ and $M^{(\be )}$ are of the same sign. So there is no nontrivial 
solution to $M=0$. This shows the failure of search for a persistent 
homoclinic orbit.

The failures of the Melnikov prediction on both a persistent heteroclinic orbit and a persistent 
homoclinic orbit indicate that there may be no chaos in the current case of perturbations.
In general, it is difficult for autonomous 
perturbations as in the current case to generate chaos. That is, DC current bias usually does 
not generate chaotic flux dynamics. Our numerical simulations 
indicate that there is no chaos under the current perturbations.

\section{The Case of $f=-\al u_t +\ga + a \sin \om t$ \label{CAF}}

First consider the ordinary differential equation (ODE) by setting $\pa_x =0$,
\[
u_{tt} = \sin u +\e \left [ -\al u_t + \ga +a \sin \om t \right ].
\]
The ($\e =0$) fixed point $u=u_t =0$ turns into a periodic orbit when $\e >0$,
\[
u_* = \e u_1 +\e^2 u_2 + \cdots .
\]
To the leading order $\O (\e )$, 
\[
u_{1tt} = u_1 + \ga +a \sin \om t
\]
which has the periodic solution
\[
u_1 = -\ga - \frac{a}{1+\om^2} \sin \om t .
\]
Under the ODE flow, the periodic orbits $u_*$ and $2\pi + u_*$ are normally hyperbolic and a Melnikov 
integral can locate a heteroclinic cycle connecting these two periodic orbits \cite{Li04e}. As a result,
there is a region in the external parameter space that supports chaos \cite{Li04e}. 

Under the PDE flow (\ref{PSG}), the stability of these periodic orbits $u_*$ and $2\pi + u_*$ is 
complicated with parametric resonances as discussed in the open problems in \cite{Li04e}. As a result of 
this complication, the existence of a heteroclinic orbit connecting the two periodic orbits and 
the existence of chaos are open rigorous mathematical problems. Consider the extended system,
\[
\left \{ \begin{array}{l} u_{tt} = c^2 u_{xx} +\sin u +\e [ -\al u_t + \ga +a \sin \th ],
\cr \dot{\th} = \om .\cr \end{array} \right.
\]
When $\e =0$, the periodic orbit $u=u_t=0$, $\th \in \mathbb{T}^1$ has two unstable eigenvalues, two stable 
eigenvalues, and the rest neutral eigenvalues. This leads to the following invariant manifold theorem.
\begin{theorem}
When $c \in (1/2,1)$ and $\e \geq 0$ is sufficiently small, each of the two periodic orbits ($u_*, \th = \om t$)
and ($2\pi + u_*, \th = \om t$) has a $2$ co-dimensional $C^m$ ($m \geq 3$) center-unstable manifold $W^{cu}_\pm$,
a $2$ co-dimensional $C^m$ center-stable manifold $W^{cs}_\pm$, and a $4$ co-dimensional $C^m$ 
center manifold $W^{c}_\pm$ in the phase space ($u,u_t,\th $) $\in H^{n+1}\times H^n \times \mathbb{T}^1$ ($n \geq 1$).
$W^{c}_\pm = W^{cu}_\pm \cap W^{cs}_\pm$. With $W^{c}_\pm$ as the base, $W^{cu}_\pm$ and $W^{cs}_\pm$ are 
fibered by $2$-dimensional Fenichel fibers.
\end{theorem}
Due to parametric resonances, dynamics inside $W^{c}_\pm$ still contains unknown number of unstable and stable 
modes with growth or decay rates of $\O (\e )$. Melnikov integrals can detect orbits in $W^{cs}_\pm$ that 
approach ($2\pi + u_*, \th = \om t$) or ($u_*, \th = \om t$) in backward time. In forward time, the destiny 
of these orbits is unknown --- This is precisely the open problem \cite{Li04e}. 

\subsection{Heteroclinic Cycle Type Connection Search}

Specifically, the Melnikov integrals measuring the simultaneous intersection between $W^{cu}_\pm$ and $W^{cs}_\mp$ are given by
\begin{eqnarray*}
M_{j\ell } &=& \int_{-\infty}^{+\infty} \int_0^{2\pi} \left \{ \frac{\pa F_\ell}{\pa u_t} 
\left [-\al u_t +\ga +a \sin \om t \right ] \right \} \bigg |_{u=u_j} dx dt \\
&=& \al M_{j\ell }^{(\al )} + \ga M_{j\ell }^{(\ga )} + a \cos (\om \rho ) M_{j\ell }^{(c)}
+ a \sin (\om \rho ) M_{j\ell }^{(s)} , 
\end{eqnarray*}
where ($j=1,2; \ \ell =0,1$) and $u_j$'s are given in (\ref{cyc1})-(\ref{cyc2}),
$M_{j\ell }^{(\cdot )}$ are functions of $c$ and $\Dl \rho = \hrho - \sg \rho$, $M_{j\ell }^{(c)}$ and 
$M_{j\ell }^{(s)}$ also depend on $\om$, and specifically  
\begin{eqnarray*}
M_{j\ell }^{(\al )} &=& - \int_{-\infty}^{+\infty} \int_0^{2\pi} \left \{ \frac{\pa F_\ell}{\pa u_t}
u_t \right \} \bigg |_{u=u_j} dx d\tau , \\
M_{j\ell }^{(\ga )} &=& \int_{-\infty}^{+\infty} \int_0^{2\pi} \left \{ \frac{\pa F_\ell}{\pa u_t}
 \right \} \bigg |_{u=u_j} dx d\tau , \\
M_{j\ell }^{(c)} &=& \int_{-\infty}^{+\infty} \int_0^{2\pi} \left \{ \frac{\pa F_\ell}{\pa u_t}
\sin (\om \tau )\right \} \bigg |_{u=u_j} dx d\tau , \\
M_{j\ell }^{(s)} &=& \int_{-\infty}^{+\infty} \int_0^{2\pi} \left \{ \frac{\pa F_\ell}{\pa u_t}
\cos (\om \tau )\right \} \bigg |_{u=u_j} dx d\tau .
\end{eqnarray*}
It turns out from numerical calculations that $M_{j\ell }^{(c)}$ and $M_{j\ell }^{(s)}$ are real and 
independent of $j$. The other integrals are the same as in previous section. Thus, the common zero of 
$M_{j\ell }$ satisfies the system
\begin{eqnarray}
& & \al M_{10}^{(\al )} + \ga M_{10}^{(\ga )} + a \cos (\om \rho ) M_{10}^{(c)} + a \sin (\om \rho ) M_{10}^{(s)}= 0, \label{acm1} \\
& & \al M_{11}^{(\al )} + \ga M_{11}^{(\ga )} + a \cos (\om \rho ) M_{11}^{(c)} + a \sin (\om \rho ) M_{11}^{(s)}= 0, \label{acm2} \\
& & \al M_{10}^{(\al )} - \ga M_{10}^{(\ga )} + a \cos (\om \rho ) M_{10}^{(c)} + a \sin (\om \rho ) M_{10}^{(s)}= 0, \label{acm3} \\
& & \al M_{11}^{(\al )} - \ga M_{11}^{(\ga )} + a \cos (\om \rho ) M_{11}^{(c)} + a \sin (\om \rho ) M_{11}^{(s)}= 0. \label{acm4}
\end{eqnarray}
There is no non-trivial solution to this system (\ref{acm1})-(\ref{acm4}), which shows a failure in searching for 
a simultaneous intersection between $W^{cu}_\pm$ and $W^{cs}_\mp$. Our numerical simulations indicate that 
there is no chaos of the type associated with heteroclinic cycles.

\subsection{Heteroclinic Orbit Type Connection Search \label{HC5}}

Next we shall search for an intersection between $W^{cu}_+$ and $W^{cs}_-$ (or $W^{cu}_-$ and $W^{cs}_+$) only, in 
which case we only need to solve equations 
(\ref{acm1})-(\ref{acm2}). Eliminating $\al$, we have 
\[
\ga A + a \cos (\om \rho ) B + a \sin (\om \rho ) C = 0 , 
\]
where
\begin{eqnarray*}
A &=& M_{10}^{(\ga )} M_{11}^{(\al )} - M_{11}^{(\ga )} M_{10}^{(\al )} , \\
B &=& M_{10}^{(c)} M_{11}^{(\al )} - M_{11}^{(c)} M_{10}^{(\al )} , \\
C &=& M_{10}^{(s)} M_{11}^{(\al )} - M_{11}^{(s)} M_{10}^{(\al )} .
\end{eqnarray*}
Thus as long as $B$ and $C$ are not simultaneously zero, there are always non-trivial solutions:
\[
\sin (\om \rho +\th_* ) = - \frac{\ga A}{a \sqrt{B^2+C^2}} , 
\]
where 
\[
\sin \th_* = \frac{B}{\sqrt{B^2+C^2}} , \quad \cos \th_* = \frac{C}{\sqrt{B^2+C^2}} .
\]
So we get the criterion that when 
\[
|a| > \chi |\ga | ,  \quad \text{where } \chi = \frac{|A|}{\sqrt{B^2+C^2}} ; 
\]
there is an intersection between $W^{cu}_+$ and $W^{cs}_-$ (or $W^{cu}_-$ and $W^{cs}_+$). The values of $\chi$ is shown in Table \ref{T5}.
\begin{table}
$$\begin{array}{|c|c|c|c|c|c|}\hline
\Delta \rho & -10 & -5 & 0 & 5 & 10\\ \hline
\omega =0.1 & 1.01 & 1.01 & 0.95 & 1.01 & 1.01\\ \hline
\omega =0.5 & 1.32 & 1.34 & 0.58 & 1.34 & 1.32\\ \hline
\omega = 1.0 & 2.51 & 2.53 & 0.67 & 2.53 & 2.51 \\ \hline
\omega =2.0 & 11.61 & 11.68 & 10.25 & 11.68 & 11.61 \\ \hline
\omega =3.0 & 55.74 & 56.00 & 53.06 & 56.00 & 55.74 \\ \hline
\omega =4.0 & 264.3 & 262.0 & 149.7 & 262.0 & 264.3 \\ \hline
\omega =5.0 & 1290. & 1231. & 507. & 1231. & 1290. \\ \hline
\end{array}$$
\caption{The table of $\chi$ for the heteroclinic orbit case in subsection \ref{HC5}, $c=0.75$.}
\label{T5}
\end{table}

\begin{figure}[ht] 
\includegraphics[width=4.0in,height=3.0in]{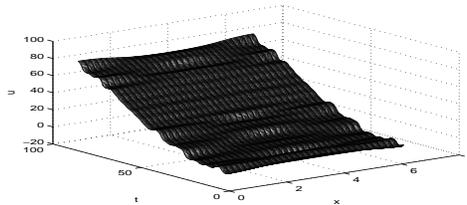}
\caption{The chaotic dynamics in subsection \ref{HC5} ($u$-plot).}
\label{fc5a}
\end{figure}

\begin{figure}[ht] 
\includegraphics[width=4.0in,height=3.0in]{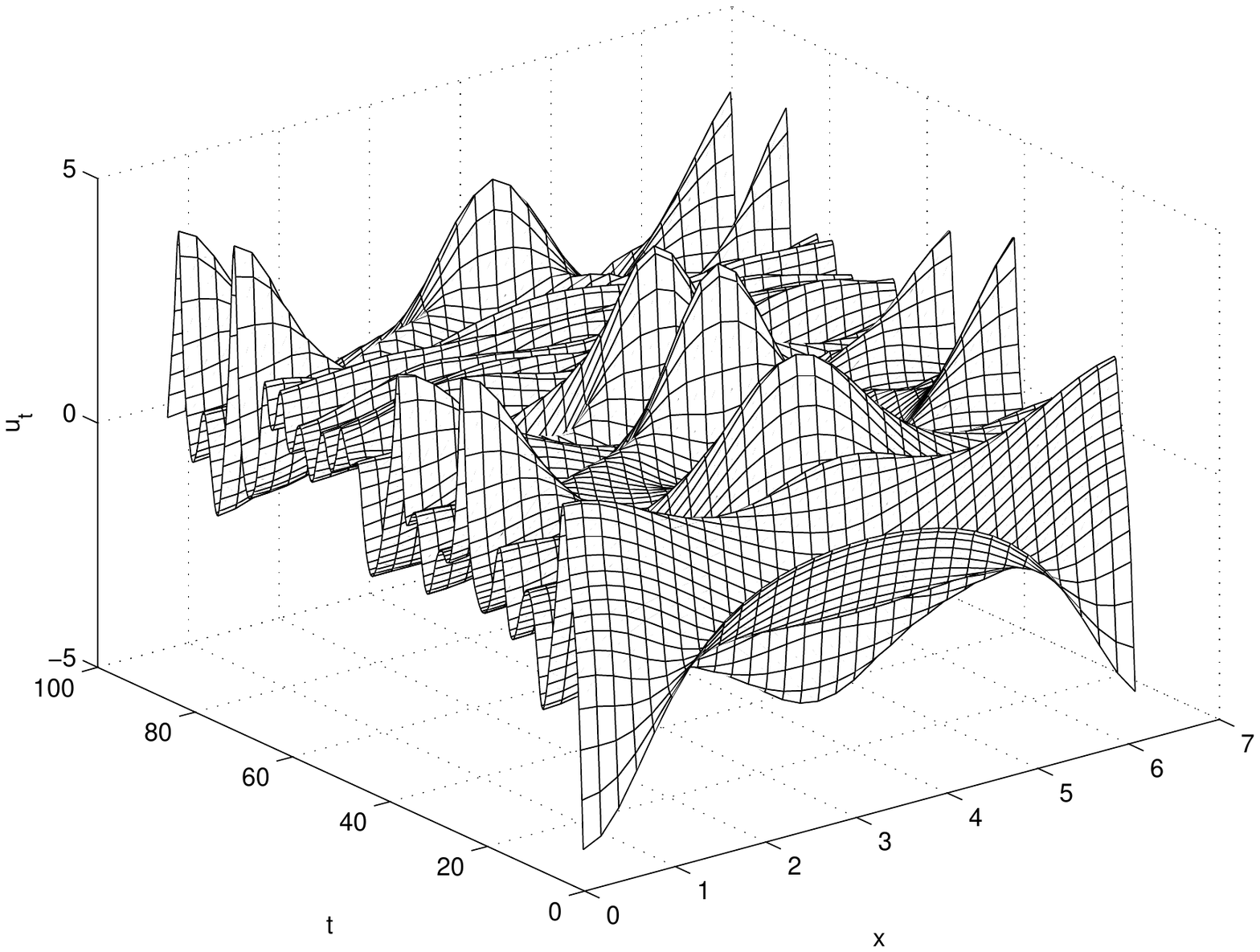}
\caption{The chaotic dynamics in subsection \ref{HC5} ($u_t$-plot).}
\label{fc5b}
\end{figure}

We conducted a numerical simulation of the dynamics with the following 
setup:
\[
\e =0.1, a=1.0, \omega =0.55, c=0.75, \alpha =0.257, \gamma =0.5;
\]
and $64$ elements are used to divide the spatial period [$0, 2\pi$], 
the time step is 1/40 of the forcing period $\frac{2\pi}{\omega}$,
finally the initial condition is given by (\ref{cyc1}) with ($t=0$, $\rho = 0$, $\hrho = 0$,
hence $\Dl \rho = 0$).
The numerical result indicates chaotic dynamics as shown in Figures \ref{fc5a}-\ref{fc5b}.
One can see a clear monotonely shifting in the $u(x,t)$ plot. This indicates that the 
chaos is due to the persistent heteroclinic orbit (the hetroclinic cycle does not persist as 
shown in last subsection). The shifting is due to the fact that when it 
ends up outside the ``eye'' in Figure \ref{topi}, the orbit will travel up to the other eye. 
If we view the range of $u(x,t)$ as a circle (i.e. mod $2\pi$), then the shifting will disappear.
The $u_t$ plot in Figure \ref{fc5b} also illustrates this effect. 

\subsection{Homoclinic Type Connection Search}

To search for the homoclinic type connection near $u_+=0$, we need to calculate the Melnikov integral (which 
measures the intersection between $W^{cu}_+$ and $W^{cs}_+$):
\begin{equation}
M = \int_{-\infty}^{+\infty} \int_0^{2\pi} \left \{ \frac{\pa F_1}{\pa u_t} 
\left [-\al u_t +\ga +a \sin \om t \right ] \right \} \bigg |_{u=u_1} dx dt, 
\label{HMI2}
\end{equation}
where $u_1$ is given in (\ref{lhd}), $\frac{\pa F_1}{\pa u_t}$ is given by (\ref{MV2}) where 
$\Phi$ is now given by (\ref{nbp}). 
\[
M = \al M^{(\al )} + \ga M^{(\ga )} + a \cos (\om \hrho /\sg ) M^{(c)}
+ a \sin (\om \hrho /\sg ) M^{(s)} , 
\]
where  
\begin{eqnarray*}
M^{(\al )} &=& - \frac{1}{\sg}\int_{-\infty}^{+\infty} \int_0^{2\pi} \left \{ \frac{\pa F_1}{\pa u_t}
u_t \right \} \bigg |_{u=u_1} dx d\htau , \\
M^{(\ga )} &=& \frac{1}{\sg}\int_{-\infty}^{+\infty} \int_0^{2\pi} \left \{ \frac{\pa F_1}{\pa u_t}
 \right \} \bigg |_{u=u_1} dx d\htau , \\
M^{(c)} &=& \frac{1}{\sg}\int_{-\infty}^{+\infty} \int_0^{2\pi} \left \{ \frac{\pa F_1}{\pa u_t}
\sin (\om \htau /\sg )\right \} \bigg |_{u=u_1} dx d\htau , \\
M^{(s)} &=& \frac{1}{\sg}\int_{-\infty}^{+\infty} \int_0^{2\pi} \left \{ \frac{\pa F_1}{\pa u_t}
\cos (\om \htau /\sg )\right \} \bigg |_{u=u_1} dx d\htau .
\end{eqnarray*}
As before, $M^{(\ga )} = 0$. It turns out that both $M^{(c)}$ and $M^{(s)}$ are zero 
(in fact $\frac{\pa F_1}{\pa u_t}_{u=u_1}$ has zero spatial mean). Thus there is no nontrivial 
solution to $M=0$. This shows a failure in searching for a persistent homoclinic orbit.
Our numerical simulations indicate that there is no chaos of the type associated with homoclinic orbits.

\section{The Case of $f=-\al u_t +\ga \cos x + a \sin \om t$ \label{ACF}}

In this case, there is no guarantee that the ($\e =0$) fixed point $u=u_t=0$ will persist into 
a periodic orbit. The existence of heteroclinic orbits and chaos is a far more open problem than the 
case in the previous section. Again consider the extended system,
\[
\left \{ \begin{array}{l} u_{tt} = c^2 u_{xx} +\sin u +\e [ -\al u_t + \ga \cos x +a \sin \th ],
\cr \dot{\th} = \om .\cr \end{array} \right.
\]
When $\e =0$, the periodic orbit $u=u_t=0$, $\th \in \mathbb{T}^1$ has two unstable eigenvalues, two stable 
eigenvalues, and the rest neutral eigenvalues. This leads to the following invariant manifold theorem.
\begin{theorem}
When $c \in (1/2,1)$ and $\e \geq 0$ is sufficiently small, there are a $2$ co-dimensional $C^m$ ($m \geq 3$) 
center-unstable manifold $W^{cu}_\pm$,
a $2$ co-dimensional $C^m$ center-stable manifold $W^{cs}_\pm$, and a $4$ co-dimensional $C^m$ 
center manifold $W^{c}_\pm$ in the neighborhoods of $u=0, 2\pi$; $u_t=0$, $\th \in \mathbb{T}^1$ (where 
$u=0$ corresponds to `$+$' and $u=2\pi$ corresponds to `$-$') in the phase space ($u,u_t,\th $) 
$\in H^{n+1}\times H^n \times \mathbb{T}^1$ ($n \geq 1$).
$W^{c}_\pm = W^{cu}_\pm \cap W^{cs}_\pm$. With $W^{c}_\pm$ as the base, $W^{cu}_\pm$ and $W^{cs}_\pm$ are 
fibered by $2$-dimensional Fenichel fibers.
\end{theorem}

\subsection{Heteroclinic Cycle Type Connection Search \label{HCC6}}

Melnikov integrals can detect orbits in $W^{cu}_\pm \cap W^{cs}_\mp$. Unfortunately both the forward 
and the backward destinies of such orbits are unknown. The Melnikov integrals are given by
\begin{eqnarray*}
M_{j\ell } &=& \int_{-\infty}^{+\infty} \int_0^{2\pi} \left \{ \frac{\pa F_\ell}{\pa u_t} 
\left [-\al u_t +\ga \cos x +a \sin \om t \right ] \right \} \bigg |_{u=u_j} dx dt \\
&=& \al M_{j\ell }^{(\al )} + \ga M_{j\ell }^{(\ga )} + a \cos (\om \rho ) M_{j\ell }^{(c)}
+ a \sin (\om \rho ) M_{j\ell }^{(s)} , 
\end{eqnarray*}
where ($j=1,2; \ \ell =0,1$) and $u_j$'s are given in (\ref{cyc1})-(\ref{cyc2}),
$M_{j\ell }^{(\cdot )}$ are functions of $c$ and $\Dl \rho = \hrho - \sg \rho$, $M_{j\ell }^{(c)}$ and 
$M_{j\ell }^{(s)}$ also depend on $\om$, and specifically  
\begin{eqnarray*}
M_{j\ell }^{(\al )} &=& - \int_{-\infty}^{+\infty} \int_0^{2\pi} \left \{ \frac{\pa F_\ell}{\pa u_t}
u_t \right \} \bigg |_{u=u_j} dx d\tau , \\
M_{j\ell }^{(\ga )} &=& \int_{-\infty}^{+\infty} \int_0^{2\pi} \left \{ \frac{\pa F_\ell}{\pa u_t}
\cos x \right \} \bigg |_{u=u_j} dx d\tau , \\
M_{j\ell }^{(c)} &=& \int_{-\infty}^{+\infty} \int_0^{2\pi} \left \{ \frac{\pa F_\ell}{\pa u_t}
\sin (\om \tau )\right \} \bigg |_{u=u_j} dx d\tau , \\
M_{j\ell }^{(s)} &=& \int_{-\infty}^{+\infty} \int_0^{2\pi} \left \{ \frac{\pa F_\ell}{\pa u_t}
\cos (\om \tau )\right \} \bigg |_{u=u_j} dx d\tau .
\end{eqnarray*}
It turns out from numerical calculations that all the integrals $M_{j\ell }^{(\al )}$, $M_{j\ell }^{(\ga )}$, $M_{j\ell }^{(c)}$ 
and $M_{j\ell }^{(s)}$ are 
real and independent of $j$. The other integrals are the same as in previous section. Thus, the common zero of 
$M_{j\ell }$ satisfies the system
\begin{eqnarray}
& & \al M_{10}^{(\al )} + \ga M_{10}^{(\ga )} + a \cos (\om \rho ) M_{10}^{(c)} + a \sin (\om \rho ) M_{10}^{(s)}= 0, \label{facm1} \\
& & \al M_{11}^{(\al )} + \ga M_{11}^{(\ga )} + a \cos (\om \rho ) M_{11}^{(c)} + a \sin (\om \rho ) M_{11}^{(s)}= 0, \label{facm2} \\
& & \al M_{10}^{(\al )} + \ga M_{10}^{(\ga )} + a \cos (\om \rho ) M_{10}^{(c)} + a \sin (\om \rho ) M_{10}^{(s)}= 0, \label{facm3} \\
& & \al M_{11}^{(\al )} + \ga M_{11}^{(\ga )} + a \cos (\om \rho ) M_{11}^{(c)} + a \sin (\om \rho ) M_{11}^{(s)}= 0. \label{facm4}
\end{eqnarray}
Notice that equations (\ref{facm3})-(\ref{facm4}) are identical with equations (\ref{facm1})-(\ref{facm2}). This 
degeneracy prohibits the application of the implicit function theorem in measuring the intersection between $W^{cu}_\pm$ and $W^{cs}_\mp$.
So we have no conclusion as whether or not $W^{cu}_\pm$ and $W^{cs}_\mp$ can simultaneously intersect.

\begin{figure}[ht] 
\includegraphics[width=4in,height=3in]{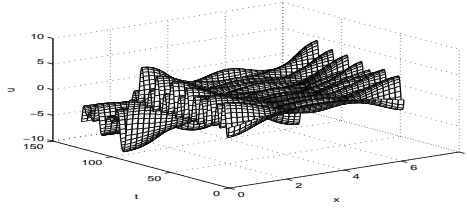}
\caption{The chaotic dynamics in subsection \ref{HCC6}.}
\label{fc6}
\end{figure}

We conducted a numerical simulation of the dynamics with the following 
setup:
\[
\e =0.1, a=5.1, \omega =1.0, c=0.75, \alpha =0.257, \gamma =3.0;
\]
and $64$ elements are used to divide the spatial period [$0, 2\pi$], 
the time step is 1/40 of the forcing period $\frac{2\pi}{\omega} = 2\pi$,
finally the initial condition is given by (\ref{cyc1}) with ($t=0$, $\rho = 0$, $\hrho = 0$,
hence $\Dl \rho = 0$).
The numerical result indicates chaotic dynamics as shown in Figure \ref{fc6}.
In fact, it seems that the chaotic dynamics loops around the eye in Figure 
\ref{topi}, i.e. around a cycle, rather than monotonely travels up  
to different eyes as in Figure \ref{fc5a}. Possible explanation is that the chaotic dynamics 
is induced by persistent heteroclinic cycles even though our Melnikov calculation above has no 
conclusion.

\subsection{Heteroclinic Orbit Type Connection Search \label{HC6}}

Next we shall search for an intersection between $W^{cu}_+$ and $W^{cs}_-$ (or $W^{cu}_-$ and $W^{cs}_+$) only, in 
which case we only need to solve equations (\ref{facm1})-(\ref{facm2}). Eliminating $\al$, we have 
\[
\ga A + a \cos (\om \rho ) B + a \sin (\om \rho ) C = 0 , 
\]
where
\begin{eqnarray*}
A &=& M_{10}^{(\ga )} M_{11}^{(\al )} - M_{11}^{(\ga )} M_{10}^{(\al )} , \\
B &=& M_{10}^{(c)} M_{11}^{(\al )} - M_{11}^{(c)} M_{10}^{(\al )} , \\
C &=& M_{10}^{(s)} M_{11}^{(\al )} - M_{11}^{(s)} M_{10}^{(\al )} .
\end{eqnarray*}
Thus as long as $B$ and $C$ are not simultaneously zero, there are always non-trivial solutions:
\[
\sin (\om \rho +\th_1 ) = - \frac{\ga A}{a \sqrt{B^2+C^2}} , 
\]
where 
\[
\sin \th_1 = \frac{B}{\sqrt{B^2+C^2}} , \quad \cos \th_1 = \frac{C}{\sqrt{B^2+C^2}} .
\]
So we get the criterion that when 
\begin{equation}
|a| > \chi |\ga | ,  \quad \text{where } \chi = \frac{|A|}{\sqrt{B^2+C^2}} ; 
\label{pa62}
\end{equation}
there is a heteroclinic orbit. The values of $\chi$ is shown in Table \ref{T6}. 
\begin{table}
$$\begin{array}{|c|c|c|c|c|c|}\hline
\Delta \rho & -10 & -5 & 0 & 5 & 10\\\hline
\omega=0.1 & 36.05 & 1.27 & 2.35 & 1.27 & 36.05\\\hline
\omega =0.5 & 47.17 & 1.67 & 1.44 & 1.67 & 47.17 \\\hline
\omega =1.0 & 89.36 & 3.16 & 1.64 & 3.16 & 89.36 \\\hline
\omega=2.0 & 413.3 & 14.59 & 25.23 & 14.59 & 413.3\\\hline
\omega =3.0 &1985. & 69.93 & 105.9 & 69.93 & 1985.\\\hline
\omega=4.0 & 9411. & 327.2 & 387.8 & 327.2 & 9411.\\\hline
\omega =5.0 & 45950. & 1538 & 1214. & 1538. & 45950.\\\hline
\end{array}$$
\caption{The table of $\chi$ for the heteroclinic orbit case in subsection \ref{HC6}, $c=0.75$. }
\label{T6}
\end{table}
The chaotic dynamics in Figure \ref{fc6} is in the parameter regime predicted by (\ref{pa62})

\subsection{Homoclinic Type Connection Search \label{HM6}}

To search for the homoclinic type connection near $u_+=0$, we need to calculate the Melnikov integral (which 
measures the intersection between $W^{cu}_+$ and $W^{cs}_+$):
\begin{equation}
M = \int_{-\infty}^{+\infty} \int_0^{2\pi} \left \{ \frac{\pa F_1}{\pa u_t} 
\left [-\al u_t +\ga \cos x +a \sin \om t \right ] \right \} \bigg |_{u=u_1} dx dt, 
\label{HMI3}
\end{equation}
where $u_1$ is given in (\ref{lhd}), $\frac{\pa F_1}{\pa u_t}$ is given by (\ref{MV2}) where 
$\Phi$ is now given by (\ref{nbp}). 
\[
M = \al M^{(\al )} + \ga M^{(\ga )} + a \cos (\om \hrho /\sg ) M^{(c)}
+ a \sin (\om \hrho /\sg ) M^{(s)} , 
\]
where  
\begin{eqnarray*}
M^{(\al )} &=& - \frac{1}{\sg}\int_{-\infty}^{+\infty} \int_0^{2\pi} \left \{ \frac{\pa F_1}{\pa u_t}
u_t \right \} \bigg |_{u=u_1} dx d\htau , \\
M^{(\ga )} &=& \frac{1}{\sg}\int_{-\infty}^{+\infty} \int_0^{2\pi} \left \{ \frac{\pa F_1}{\pa u_t} \cos x
 \right \} \bigg |_{u=u_1} dx d\htau , \\
M^{(c)} &=& \frac{1}{\sg}\int_{-\infty}^{+\infty} \int_0^{2\pi} \left \{ \frac{\pa F_1}{\pa u_t}
\sin (\om \htau /\sg )\right \} \bigg |_{u=u_1} dx d\htau , \\
M^{(s)} &=& \frac{1}{\sg}\int_{-\infty}^{+\infty} \int_0^{2\pi} \left \{ \frac{\pa F_1}{\pa u_t}
\cos (\om \htau /\sg )\right \} \bigg |_{u=u_1} dx d\htau .
\end{eqnarray*}
As before, $M^{(c)}$ and $M^{(s)}$ are zero, but now $M^{(\ga )}$ is not zero.
So $M=0$ implies that 
\[
\ga = \chi \al , \text{ where } \chi = -M^{(\al )} / M^{(\ga )} .
\]
When $c = 0.75$, we found that
\[
\chi = - 6762.7.
\]
That is, comparing to the shunt loss amplitude $\al$, the AC current field amplitude $\ga$ needs 
to be quite large to generate possible homoclinic chaotic current (CC). 
\begin{figure}[ht] 
\includegraphics[width=4in,height=3in]{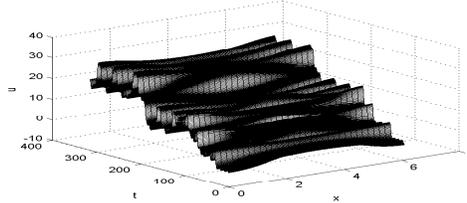}
\caption{The chaotic dynamics in subsection \ref{HM6}.}
\label{fhm6}
\end{figure}
We conducted a numerical simulation of the dynamics with the following 
setup:
\[
\e =0.01, a=0, \omega =0.55, c=0.75, \alpha =0.001, \gamma = -6.7627;
\]
and $32$ elements are used to divide the spatial period [$0, 2\pi$], 
the time step is $1/80$ of the forcing period $\frac{2\pi}{\omega}$,
finally the initial condition is given by (\ref{cyc1}) with ($t=0$, $\rho = 0$, $\hrho = 0$,
hence $\Dl \rho = 0$).
The numerical result indicates chaotic dynamics as shown in Figure \ref{fhm6}.
This chaos is still jumping around the heteroclinic loops, nevertheless most of the time 
it loops very close to the homoclinic orbits.

\section{The Components of the Global Attractor and the Bifurcation in the Perturbation Parameter \label{Bif}}

Here we study the case of $f=-\al u_t +\ga  + a \sin \om t$ in (\ref{PSG}) 
which was also studied in section \ref{CAF}. The setup is as follows:
\begin{equation}
a=1.0, \omega =0.55, c=0.75, \alpha =0.257, \gamma =0.5;
\label{cpc2}
\end{equation}
and $64$ elements are used to divide the spatial period [$0, 2\pi$], 
the time step is $1/40$ of the forcing period $\frac{2\pi}{\omega}$.

\begin{figure}[ht]
\centering
\subfigure[]{\includegraphics[width=2.3in,height=2.3in]{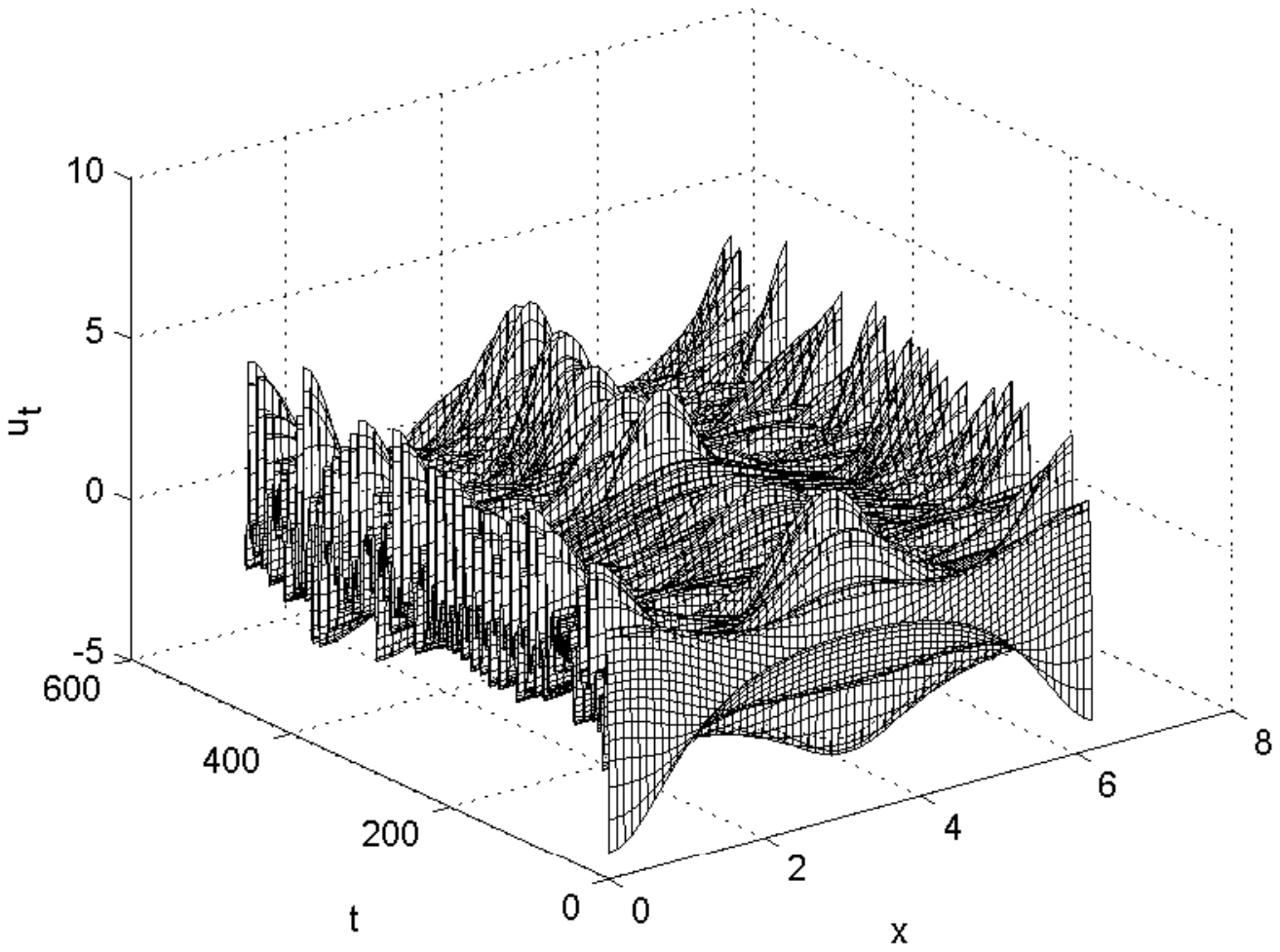}}
\subfigure[]{\includegraphics[width=2.3in,height=2.3in]{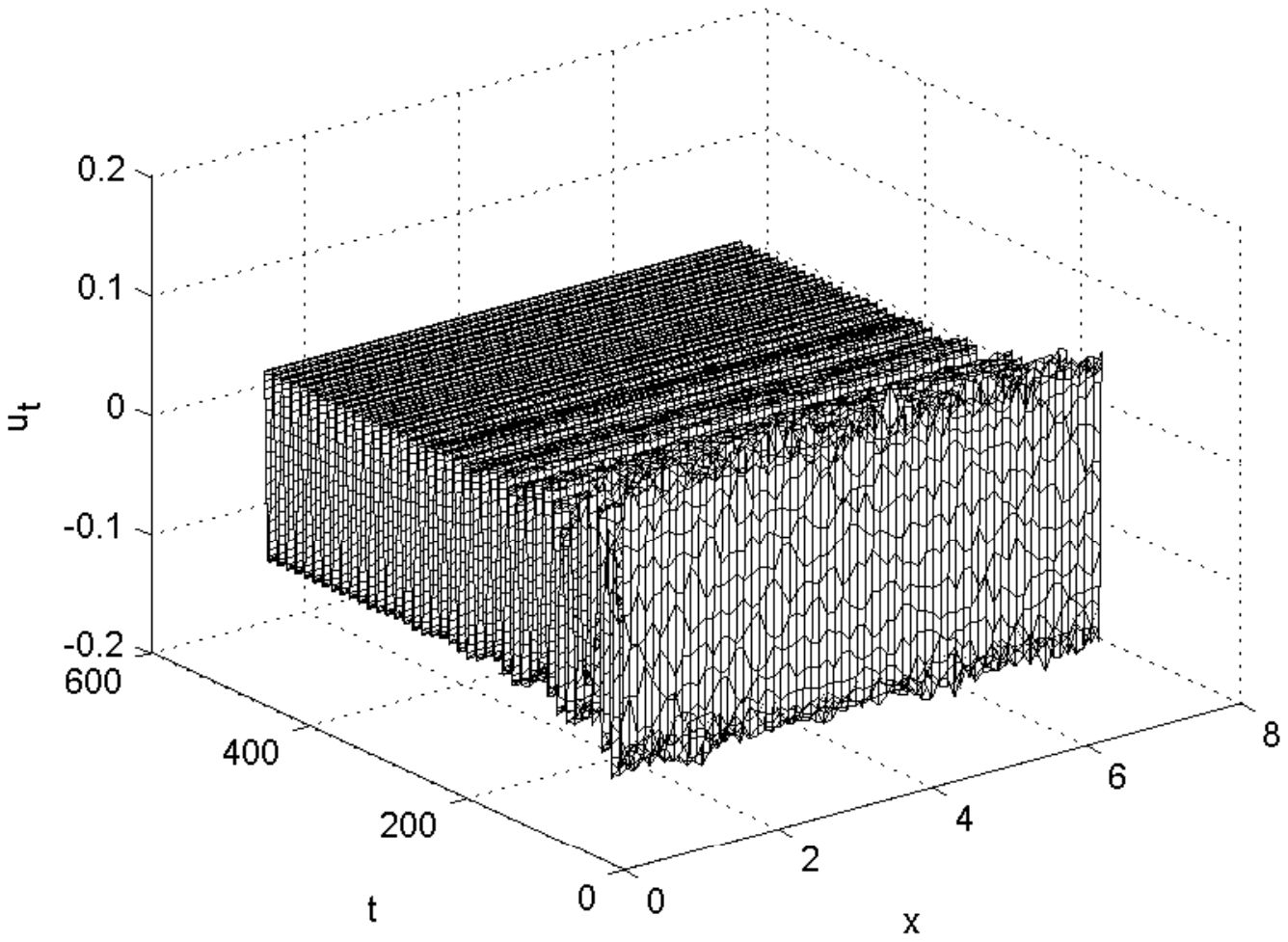}}
\caption{Two local attractors inside the global attractor. (a) is a chaotic 
attractor, and (b) is a periodic attractor.}
\label{Com}
\end{figure}

\begin{figure}[ht] 
\includegraphics[width=4in,height=3in]{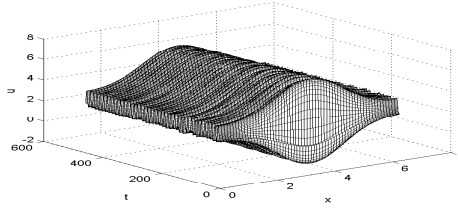}
\caption{The breather attractor $B_+$ in Table \ref{cbd}.}
\label{brea}
\end{figure}

The perturbed sine-Gordon system (\ref{PSG}) has a global attractor 
up to the translation $u \ra u +2\pi$ \cite{CV02}. Here we are interested in 
the detailed structure of the global attractor. In particular, we are interested
in the local attractors inside the global attractor. Components other than the 
local attractors, often can only attract measure zero sets of initial conditions. Figure \ref{Com} shows two local attractors inside the global attractor when 
$\e =0.1$. 
For the initial condition ($u^*, u^*_t$) 
given by (\ref{cyc1}) with $t=0$, $\rho = 0$, $\hrho = 0$ (hence $\Dl \rho = 0$,
and $u^* = \pi$); the solution approaches the chaotic 
attractor Figure \ref{Com}(a). For the initial condition 
\[
(\hu , \hu_t) = (\pi ,0.1+0.01 r(x))
\]
where $r(x)$ is a random function depending on $x$; the solution approaches the 
periodic attractor Figure \ref{Com}(b).
To explore the initial conditions more systematically, we use the following 
homotopic initial conditions
\[
(u , u_t) = (1-A) (u^*, u^*_t) + A (\tu , \tu_t)
\]
where $A \in [0,1]$ and $(\tu , \tu_t) = (\pi ,0.1)$
i.e. we dropped the random perturbation in ($\hu , \hu_t$), which does not seems 
to affect the final attractor. Another topic that we are interested in is the 
bifurcation in the perturbation parameter $\e$. By combining the two parameters 
$\e$ and $A$, we obtain the component-bifurcation diagram (Table \ref{cbd}), 
where H is the heteroclinic orbit given by (\ref{cyc1}), Q is the quasiperiodic 
solution of the integrable sine-Gordon equation ($\e =0$). U is spatially uniform 
and temporally periodic attractor as depicted in Figure \ref{Com}(b). It is 
originated from the integrable steady state $(u, u_t) = (\pi , 0)$ and modulated 
by the temporally periodic forcing $a\sin \om t$. $B_+$ is a breather attractor 
as depicted in Figure \ref{brea}. It is originated from the center of a loop of 
one of the small figure-eights in Figure \ref{topi} and modulated by the 
perturbation.
$B_-$ is a breather attractor similar to $B_+$ except that the hump is located 
at the spatial periodic boundary. It is orginated from the center of the other loop
of the small figure-eight. C is the chaotic attractor as depicted in 
Figure \ref{Com}(a). When the value of $\e$ or $A$ is larger than those in Table 
\ref{cbd}, the attractor is always U. From Table \ref{cbd}, one can see that there 
is no clear boundary between chaos attractor and regular attractor in the 
($\e, A$)-plane. The chaotic attractors appear and disappear in an irregular fashion 
when the parameters are varying. In the perturbation parameter $\e$, the bifurcation 
does not follow any simple bifurcation paradigm in low dimensional systems.
\begin{table}
\centering
\vspace{0.25in}
\begin{tabular}{|c|c|c|c|c|c|}
\hline
$\e$   & $A=0.00$ & $A=0.05$ & $A=0.10$ & $A=0.15$ & $A=0.20$  \\
\hline
$0$    &   H   &  Q       &   Q      &  Q       &   Q       \\
\hline
$0.01$ &   U   &  $B_-$   &   U      &  U       &   U       \\
\hline
$0.02$ &   U   &  U       &   U      &  U       &   U       \\
\hline
$0.03$ & $B_+$ &  U       &   U      &  U       &   U       \\
\hline
$0.04$ & C     &  C       &   C      &  U       &   U       \\
\hline
$0.05$ & C     &  C       &   U      &  U       &   U       \\
\hline
$0.06$ & U     &  C       &   U      &  U       &   U       \\
\hline
$0.07$ & C     &  C       &   U      &  U       &   U       \\
\hline
$0.08$ & C     &  U       &   C      &  U       &   U       \\
\hline
$0.09$ & U     &  C       &   C      &  U       &   U       \\
\hline
$0.1$  & C     &  U       &   U      &  U       &   U       \\
\hline
\end{tabular}
\vspace{0.5in}
\caption{The component-bifurcation diagram in section \ref{Bif}. $H$: heteroclinic orbit,
$Q$: quasiperiodic orbit, $U$: spatially uniform and temporally periodic attractor, $B_+,B_-$: 
breather attractors, $C$: chaotic attractor.}
\label{cbd}
\end{table}

\section{Ratchet Effects}

The long Josephson vortex ratchet effect refers to the effect of nonzero 
temporal average of the voltage output $u_t$ when the temporal part of the 
input forcing $f$ (\ref{PSG}) has a zero temporal average. There has
been a lot of recent interest on the Josephson vortex ratchet, see \cite{Bec05}
and the references therein. Here we will study 
the following form of the input forcing $f$ (\ref{PSG}):
\begin{equation}
f=-\al u_t +g(x) + a \sin \om t ,
\label{REF}
\end{equation}
where $g(x)$ is a spatially periodic current field of zero mean. It turns 
out that the existence of the ratchet effect depends on whether or not the 
spatial potential $G$ ($G_x = g$) is symmetric as observed experimentally \cite{Bec05}. 

\begin{figure}[ht]
\centering
\subfigure[]{\includegraphics[width=2.3in,height=2.3in]{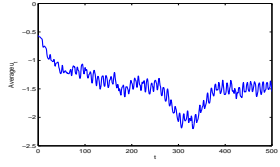}}
\subfigure[]{\includegraphics[width=2.3in,height=2.3in]{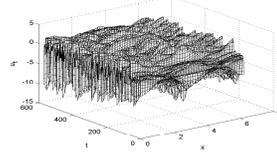}}
\subfigure[]{\includegraphics[width=2.3in,height=2.3in]{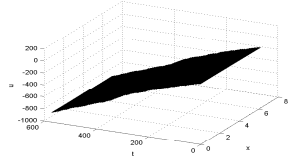}}
\caption{The temporal average of $u_t$ at $x=0$ is shown in (a), 
while (b) and (c) are spatial-temporal profiles. This is the ratchet 
effect in the asymmetric potential case.}
\label{RE}
\end{figure}

First we study the asymmetric potential case:
\[
g(x) = \ga \ \ \text{for } |x-\pi | \leq \frac{\pi}{48}; \quad -\frac{\ga}{47} 
\ \ \text{for other } x \in [0, 2\pi ].
\]
Figure \ref{RE} corresponds to the following setup:
\begin{equation}
\e =0.1, a=2.0, \omega =0.55, c=0.75, \alpha =0.14, \gamma = 100;
\label{rsu}
\end{equation}
and $64$ elements are used to divide the spatial period [$0, 2\pi$], 
the time step is $1/40$ of the forcing period $\frac{2\pi}{\omega}$,
finally the initial condition is given by (\ref{cyc1}) with 
($t=0$, $\rho = 0$, $\hrho = 0$, hence $\Dl \rho = 0$); the temporal
average of $u_t$ is done over 256 data points, i.e. 6.4 [256/40] times 
of the forcing period $\frac{2\pi}{\omega}$. One can see a clear 
nonzero temporal average of the voltage $u_t$ at $x=0$. That is, there 
is a clear ratchet effect. The spatio-temporal profile of $u_t$ is clearly
chaotic in time, while the spatio-temporal profile of $u$ is also chaotic
and dominated by drifting in time. 

\begin{figure}[ht]
\centering
\subfigure[]{\includegraphics[width=2.3in,height=2.3in]{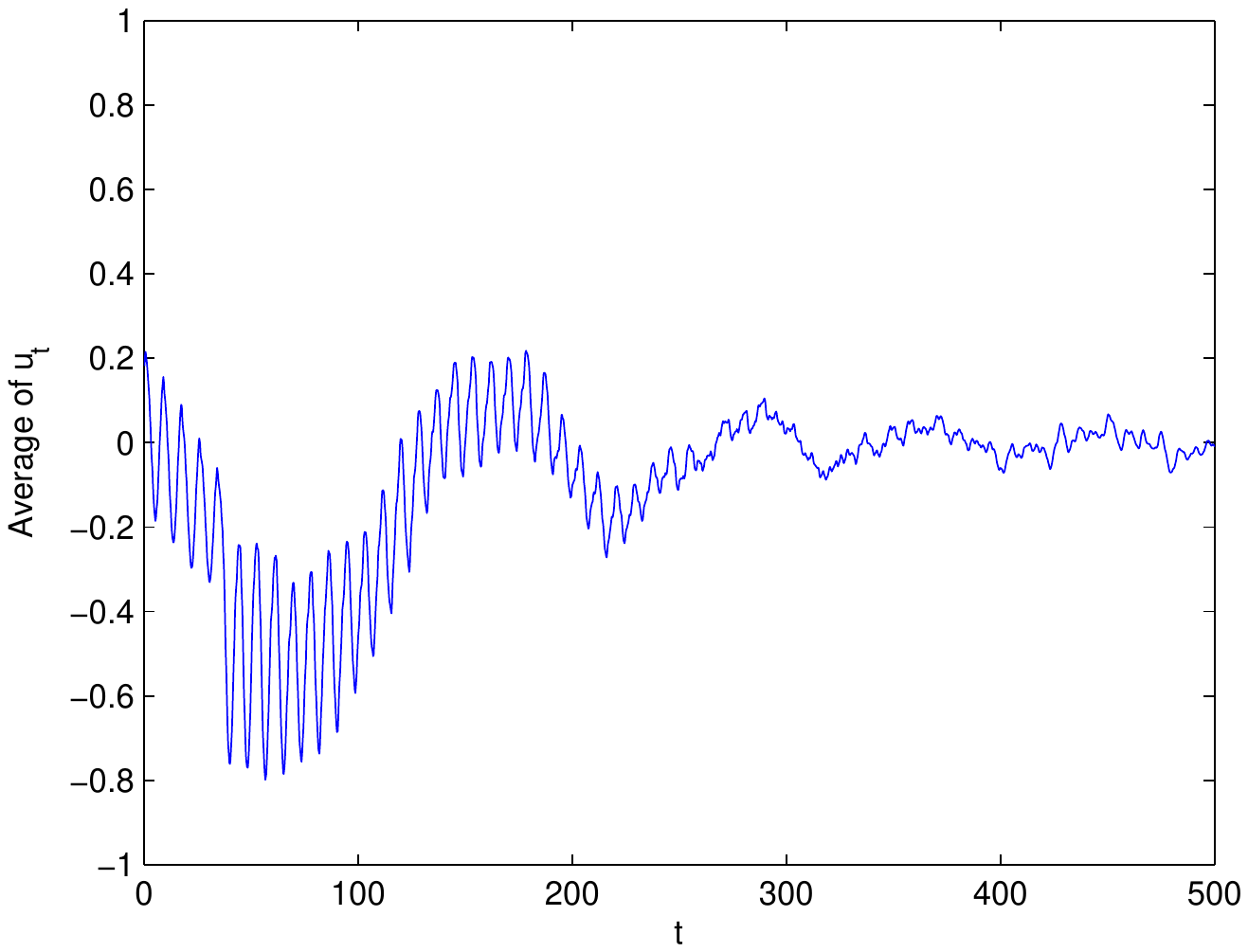}}
\subfigure[]{\includegraphics[width=2.3in,height=2.3in]{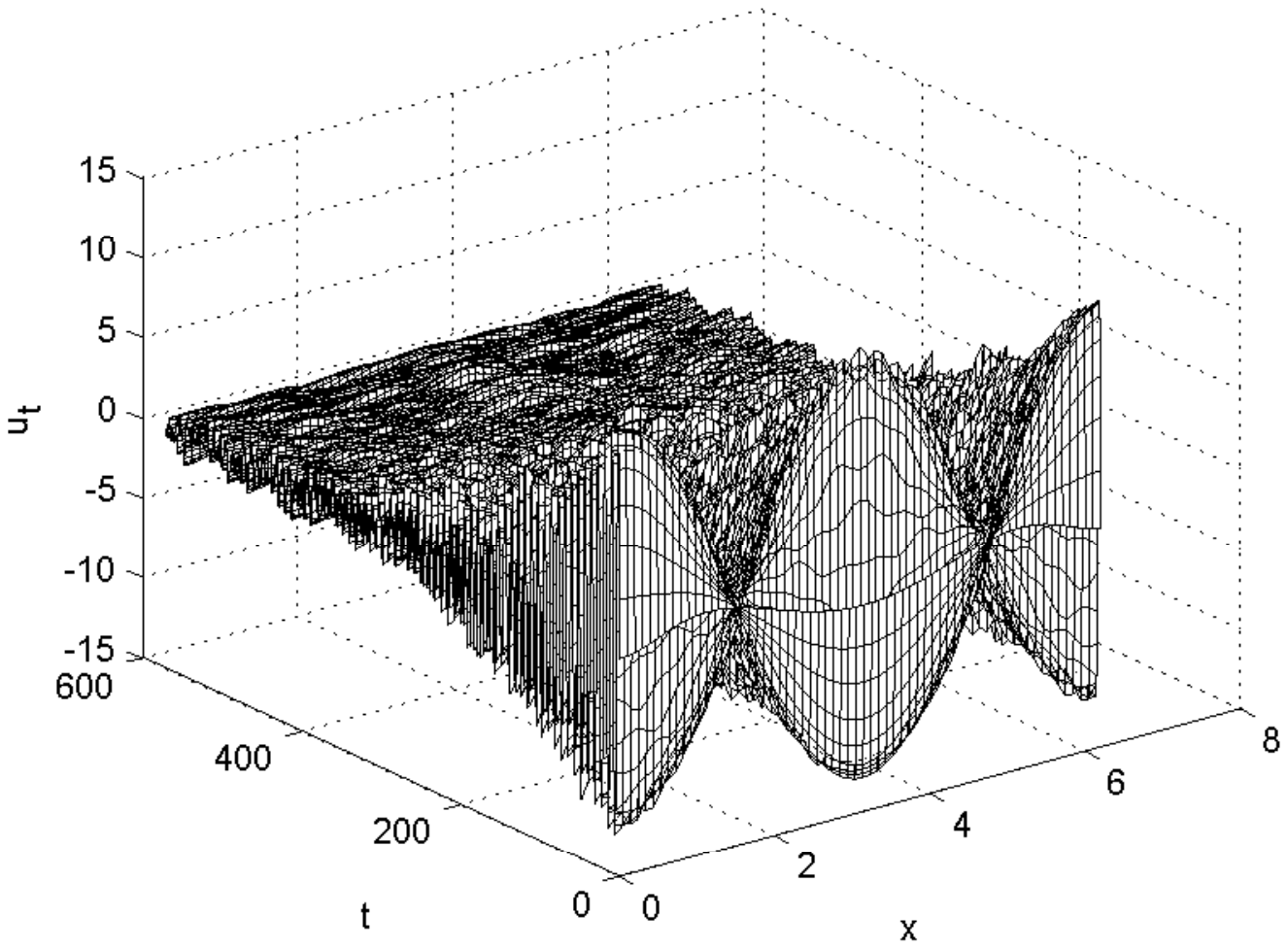}}
\subfigure[]{\includegraphics[width=2.3in,height=2.3in]{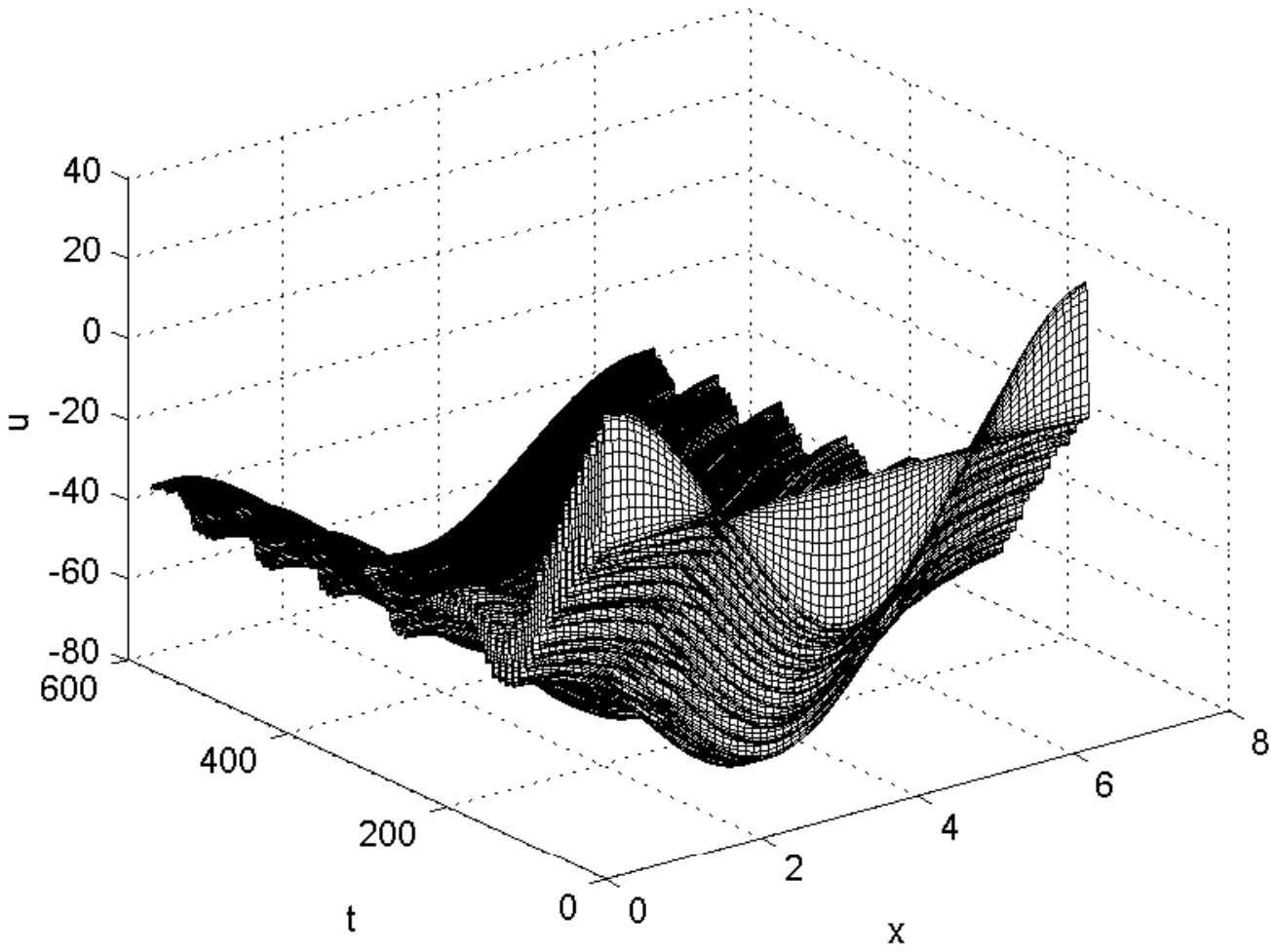}}
\caption{The temporal average of $u_t$ at $x=0$ is shown in (a), 
while (b) and (c) are spatial-temporal profiles. This corresponds to no ratchet effect
in the symmetric potential case.}
\label{NRE1}
\end{figure}

\begin{figure}[ht]
\centering
\subfigure[]{\includegraphics[width=2.3in,height=2.3in]{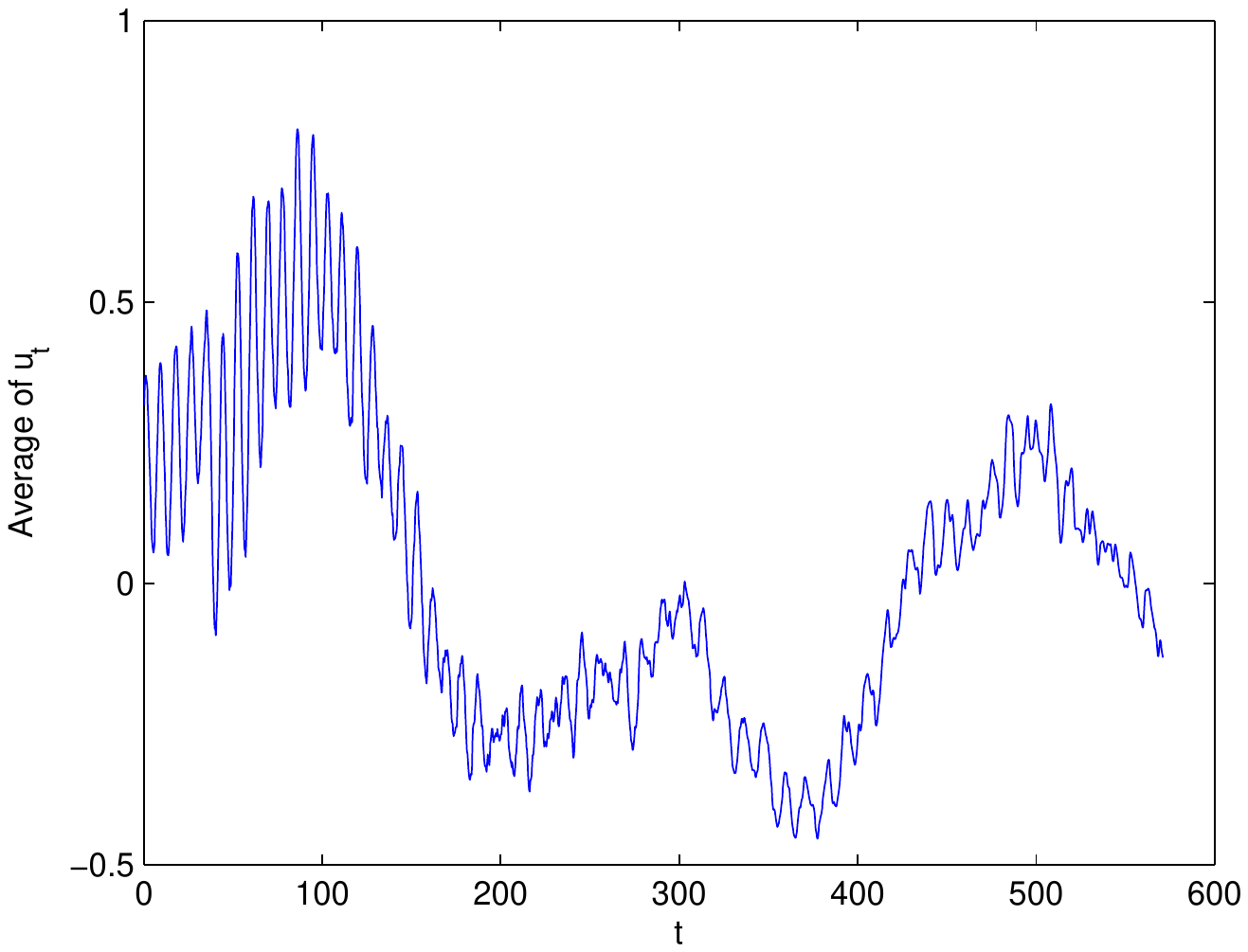}}
\subfigure[]{\includegraphics[width=2.3in,height=2.3in]{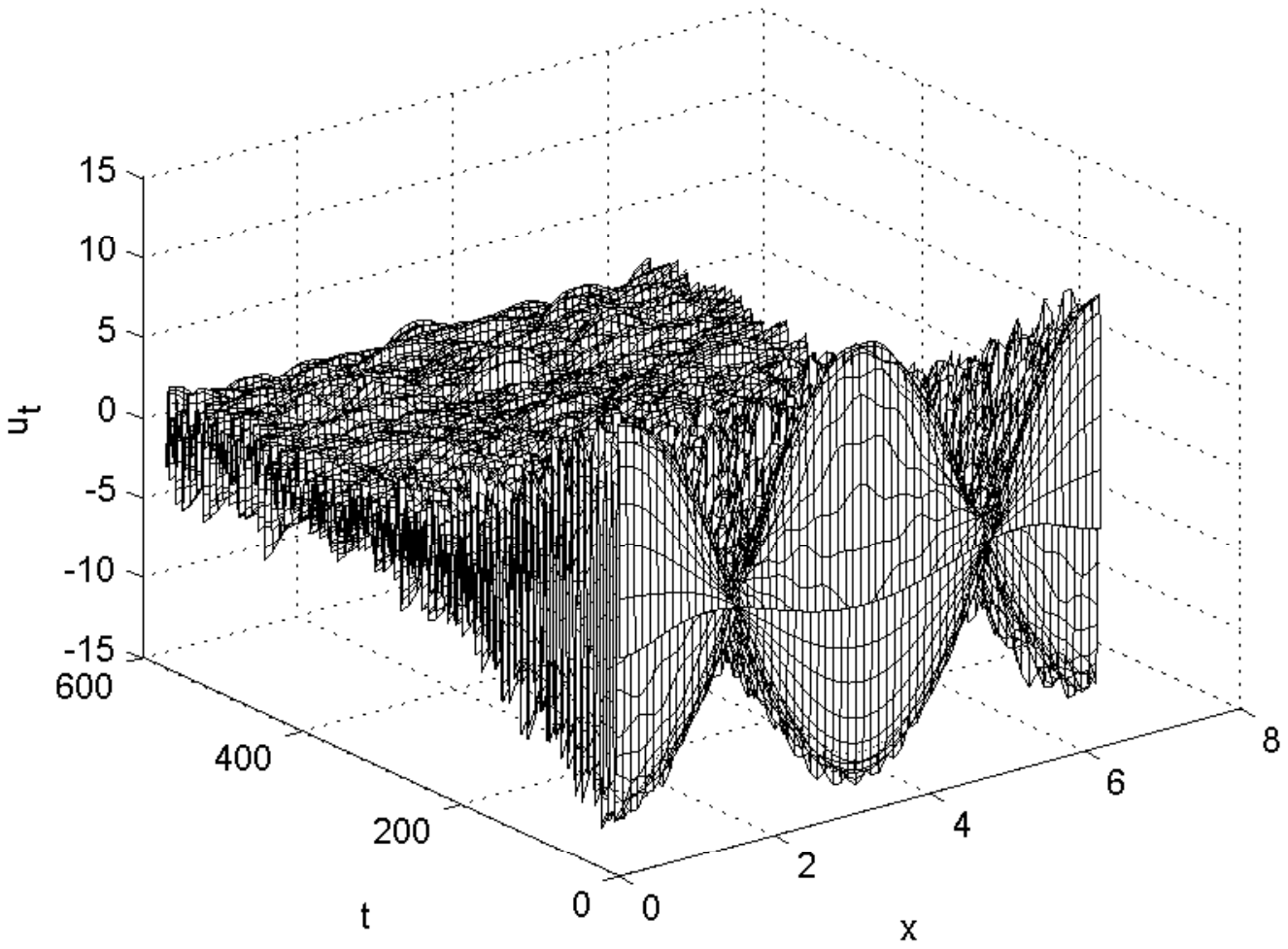}}
\subfigure[]{\includegraphics[width=2.3in,height=2.3in]{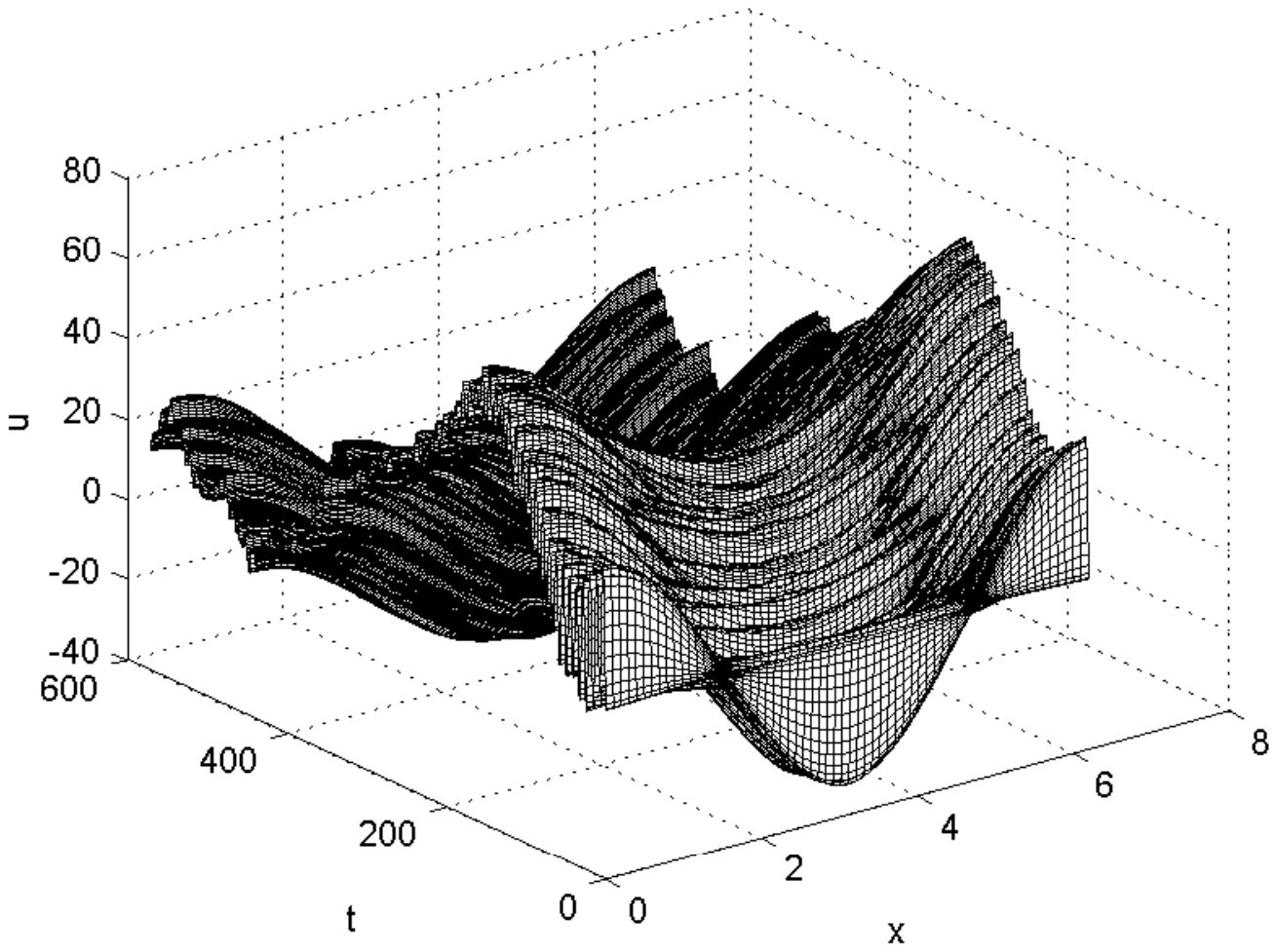}}
\caption{The temporal average of $u_t$ at $x=0$ is shown in (a), 
while (b) and (c) are spatial-temporal profiles. This also corresponds to no ratchet effect
in the symmetric potential case. The difference from Figure \ref{NRE1} is that the value 
of the parameter $a$ is different.}
\label{NRE2}
\end{figure}

Next we study the symmetric potential case: $g(x) = \ga \cos x$ 
which is the case studied in Section \ref{ACF}. Figure \ref{NRE1} 
corresponds to the same setup as (\ref{rsu}) and the rest. One can 
see that the temporal average of the voltage $u_t$ at $x=0$ is approaching 
zero in time.That is, there 
is no ratchet effect in long term. The spatio-temporal profile of $u_t$ is transiently
chaotic in time, while the spatio-temporal profile of $u$ is also transiently
chaotic in time and its drifting in time is mild. 
Figure \ref{NRE2} corresponds to the same setup as (\ref{rsu}) and 
the rest except $a=5.0$. One can 
see that the temporal average of the voltage $u_t$ at $x=0$ is also 
slowly approaching 
zero in time. That is, there 
is no ratchet effect in long term. The spatio-temporal profile of $u_t$ is 
transiently chaotic in time with a longer term, while the 
spatio-temporal profile of $u$ is also transiently
chaotic in time and it almost has no drifting in time. 

\section{Conclusion and Discussion}

Via a combination of numerical and Melnikov integral studies, we find that DC current 
bias cannot induce chaotic flux dynamics, 
while AC current bias can. The existence of a common root to the Melnikov integrals 
is a necessary condition for the existence of chaotic flux dynamics. The global attractor 
can contain co-existing local attractors e.g. a local 
chaotic attractor and a local regular attractor. In the infinite dimensional phase 
space setting, the bifurcation is very complicated. Chaotic attractors can appear and 
disappear in a random fashion. In the parameter space, there is no clear regular boundary 
between local chaotic attractors and local regular attractors.
Three types of attractors (chaos, breather, spatially 
uniform and temporally periodic attractor) are identified. Ratchet effect can be achieved 
by a current bias field which corresponds to 
an asymmetric potential as observed in experiments \cite{Bec05}, in which case the flux 
dynamics is ever lasting chaotic. 
When the current bias field corresponds to a symmetric potential, the flux dynamics 
is often transiently chaotic, in which case the ratchet effect disappears after 
sufficiently long time. 

Due to its infinite dimensionality, numerically exploring the entire phase space is 
impossible. Here we focus upon an interesting neighborhood. It is entirely possible that 
other novel structures are hidden somewhere else in the phase space. Also due to the 
infinite dimensionality, the link between existence of chaos (and homoclinic orbit) and 
existence of a common root to the Melnikov integrals becomes weaker. Finally, chaos in 
the infinite diemsional phase space is often transient. After sufficiently long time, 
the seemingly chaotic dynamics may converge to a regular attractor. 

Inside the global attractor, there may be many invariant components. An interesting 
topic is to classify these components. Of particular interest are those components which 
are local attractors. Some local attractors may be chaotic, while others may be regular. 
Different initial conditions may lead to different local attractors. A complete 
classification of all these local attractors is very challenging, especially in the 
infinite dimensional setting. This article only explores a selected set of initial 
conditions. When the value of the perturbation parameter changes, the dynamics undergoes 
bifurcations. Unlike in low dimensional systems, bifurcations in the infinite dimensional 
system are much more complicated.

\end{document}